\documentclass[preprint,12pt]{elsarticle}
\usepackage{listings}
\usepackage{verbatim}
\usepackage{epsfig}

\newcommand{\pdt}[1]{\frac{\partial #1}{\partial t}}
\newcommand{\pddt}[1]{\frac{\partial^2 #1}{\partial t^2}}

\newcommand{\pdtau}[1]{\frac{\partial #1}{\partial \tau}}
\newcommand{\pddtau}[1]{\frac{\partial^2 #1}{\partial \tau^2}}
\newcommand{\eqref}[1]{(\ref{#1})}

\title{Impact of magnetic fields on polaron dynamics in low-dimensional systems
}

\author[1]{Larissa Brizhik\fnref{fn1}}
\ead{brizhik@bitp.kyiv.ua}
\author[2]{B.M.A.G. Piette\fnref{fn2}\corref{cor1}}
\ead{b.m.a.g.piette@durham.ac.uk}

\affiliation[1]{organization={Bogolyubov Institute for Theoretical Physics of the National Academy of Sciences of Ukraine},
  addressline={Metrologichna Str., 14-b},
  postcode={03143},
  city={Kyiv},
  country={Ukraine}}

\affiliation[2]{organization={Department of Mathematical Sciences, Durham University},
  addressline={Stockton Road},
  postcode={DH1 3LE},
  city={Durham},
  country={United Kingdom}}
\fntext[fn1]{brizhik@bitp.kyiv.ua}
\fntext[fn2]{b.m.a.g.piette@durham.ac.uk}
\cortext[cor1]{Corresponding author}

\begin{document}

\begin{abstract}
  We study the impact of an external magnetic field on the long-range electron
  transport in quasi-one-dimensional materials, such as polypeptides,
  (semi)\-con\-ducting polymers and macromolecules, taking into account the
  electron-lattice interaction. At relatively strong electron-lattice
  interaction extra electrons get self-trapped in the deformation potential
  well and form stable bound states, called large polarons which in the continuum
  approximation are known as solitons.
  Here we do not use the continuum approximation but solve the system of discrete
  nonlinear equations numerically.
  We show that the impact of a magnetic field
  on polaron dynamics  depends not only on the field strength, but also on the
  parameter values of the system which define the properties of solitons such as
  their energy, amplitude and width of localisation. We also study the impact
  of a magnetic field on a polaron created by a donor complex on a chain.	
\end{abstract}
 
 \maketitle	
 
 
 \vspace{5mm}
 

\section{Introduction}
Recent advancements in nanotechnologies based on nanomaterials and
nanostructures
have stimulated the synthesis of novel low-dimensional (LD) nanomaterials with
physical and chemical properties qualitatively different from  those of other
bulk materials. This has remarkably expanded the applications of LD compounds in
numerous nanotechnologies. In particular, quasi-one-dimensional (Q1D) systems
are basic units of nanorods, nanowires, and nanobelts. LD materials provide a
directed path for charge carrier transport in Acceptor -- Molecular Chain --
Donor systems, which can occur on macroscopic distances even at room
temperatures \cite{Ing,Amdursky,Champion}. Among such Q1D compounds the most
promising ones seem to be not only conducting polymers (CPs), but also
biological macromolecules, such as polypeptides (PPs) and DNA, used in
bio-nanotechnologies and biomimicring technologies. 

There is a large class of such novel LD functional materials (see, e.g.,
\cite{Ban,Karunasena} and references therein) which enable high-performance,
low-cost optoelectronic and thermoelectric devices, which are used for energy
storage as well as in solar cell technologies, photonics, sensors, catalysis,
coatings, etc. \cite{Jia,Bhat,Poudel,Huang,Chang-solar-cells}.
This raises the question of how one can explain how the highly effective conductivity
of LD compounds can be stable in a broad range of physical conditions,
including room and higher
temperatures (see, e.g., \cite{Venka}). In reply to the statement
"The  understanding of conduction mechanisms remains
incomplete even after several decades of intensive investigation" \cite{Ugur},
we point
out that the best candidate for such a mechanism is provided by solitons, a
special type of large polarons that correspond to nonlinear electron states in
molecular chains, self-trapped in the deformation well due to the polaron effect
\cite{Davydov,BPZ-DChA}. Such solitons in physical systems and biological
macromolecules provide a long-range energy and charge transport with
almost no energy
dissipation. This is due to the compensation of the dispersive and nonlinear
effects
arising from the electron exchange interaction between neighbour sites
and the electron-phonon coupling, respectively. At intermediate values of the
electron-lattice coupling, solitons are extended over several lattice sites
and propagate along the system as a coherent running wave. At higher
concentrations, the electrons get self-trapped in a collective periodic
many-soliton state similar to a charge density wave \cite{LB-Dav-periodic}.
This theoretical result is supported by the fact that "There is a strong
tendency for collective charge motion along the ordering direction of the
polymer chains" \cite{Ugur,Sirringhaus}. 

Devices based on LD materials, are often exposed to magnetic fields (MFs)
in physical conditions. Moreover, low-intensity permanent and oscillating
MFs are used in non-invasive therapies. High-intensity MFs are used in
diagnostic devices, such as tomography and so on. Therefore, it is important
to know the impact of MFs on charge transport properties of LD materials
and macromolecules. Analytically this
problem has been investigated in \cite{LB-MF-SChFr} within the continuum
approximation using a nonlinear perturbation method. Meanwhile it is
known that the discreteness of molecular chains results in the formation of the
so-called periodic Peierls-Nabarro potential relief which affects the
soliton dynamics
\cite{BEC-Peierls} and which can be critical in the presence of a MF.
Therefore, here we study the impact of a MF on electron transport in LD
systems using the discrete system of nonlinear equations for the electron
wave function and lattice displacements, taking into account the
electron-lattice interaction. For our present purposes it is not necessary
to study the interaction between the self-trapped electrons, so we will use
the term "polaron" as a more general term than "soliton". 
We consider a lattice monomer as a unit site and take into account longitudinal
displacements of the sites from their equilibrium positions caused by the
presence of an extra electron and electron-phonon coupling. This model is
described in Section \ref{Model}. The system of discrete equations does not
admit exact analytical solution, so we solve them numerically. In
Section \ref{param} we discuss the known range of parameter values for
CPs and macromolecules.
In Section \ref{donor} we derive the equations for the Donor-Polymer system.
For the numerical simulations,
described in Section \ref{num} we use 3 sets of parameters corresponding
respectively to an Amide-I vibration in PPs, an extra electron
in a PP and a CP.
Finally, in Section \ref{Concl} we summarise the obtained results,
compare them with the analytical study and discuss their implications and
further generalisation of the model.

\section{Hamiltonian of the system and equations of motion}\label{Model}

If we consider a one-dimensional molecular chain with an extra electron
in it and take into account the electron-lattice interaction, in the absence
of a  magnetic field this system is described  by the Fr\"ohlich-type
Hamiltonian
\begin{equation}
	H={{H}_{el}}+{{H}_{latt}}+{{H}_{int}},
	\label{FH}
\end{equation}
where the terms ${H}_{el}$,\, ${H}_{latt}$ and ${H}_{int}$ are the Hamiltonians
of the electron, lattice vibrations (phonons) and electron-lattice interaction:
\begin{equation}
H_{el}=\sum_{n}\left[ E_{0}\Psi_{n}^{*}\Psi_{n}-J\left(\Psi_{n}^{*}\Psi_{n+1}+\Psi_{n+1}^{*}\Psi_{n}\right) \right],
	\label{Hel}
\end{equation}
\begin{equation}
	H_{latt}=\frac{1}{2}\sum_{n}\left[ M\left(\frac{dU_{n}}{dt}\right)^{2}+\bar{W}\left(U_{n+1}-U_{n}\right)^{2}\right]
		\label{Hlatt},
	\end{equation}
\begin{equation}
	H_{int}=\sum_{n} \bar{\sigma} \Psi_{n}^{*}\Psi_{n}\left(U_{n+1}-U_{n-1}\right).		
		\label{Hint}
\end{equation}             
Here $\Psi_{n}=\Psi_{n}(t) $ is the normalised electron wave function on the $n$-th site,
$U_n=U_n(t)$
is the displacement of the $n$-th unit cell from its equilibrium position.
Moreover $E_0$ is the on-site electron energy, $M$ is the unit cell mass,
$\bar{W}$ is the elasticity of the bond between the nearest sites,
$\bar{\sigma}$ is the
electron-lattice coupling constant and  $J$ is the nearest neighbour
electron exchange energy: 
\begin{equation}
J=\frac{\hbar ^2}{2ma^2}
		\label{J}
\end{equation}
where $a$ is the lattice constant.

From this Hamiltonian one can derive the
nonlinear system of equations for the electron wave function  and lattice
displacements
\begin{eqnarray}
	&&i\hbar \pdt{\Psi_n}=E_0\Psi_{n} -J\left(\Psi_{n+1}+\Psi_{n-1}\right)
	+\bar{\sigma}\left(U_{n+1}-U_{n-1}\right)\Psi_n,
\nonumber\\
	&&M \pddt{U_n} = \bar{W}(U_{n+1}-2U_{n}+U_{n-1}) +\bar{\sigma}(|\Psi_{n+1}|^2-|\Psi_{n-1}|^2)
	\label{eq_elphB1}
\end{eqnarray}
which in the continuum approximation can be reduced to the nonlinear
Schr\"odinger
equation with the well-known soliton solution \cite{Davydov}.

In the presence of a MF we have to generalise the model to a three-dimensional
case and use the Peierls substitution for the electron momentum ${\vec p}({\vec r})$
(see \cite{LB-MF-SChFr})
\begin{equation}
	\label{mom}
	{\vec p}({\vec r}) \rightarrow	{\vec P}({\vec r})\equiv  {\vec p}({\vec r}) - e{\vec A} 
\end{equation}
where  ${\vec A}$ is the MF vector-potential,  ${\vec B}={\rm{rot}} \vec{{A}}$,
and $e$ is the  electron charge. We can represent the electron wave function as
the product 
\begin{equation}
	\label{psi3D}
	\Psi (\vec{r}, t)=\Psi _n(t)\Psi_{tr} (y,z,t)
\end{equation}
where $\Psi_{tr} (y,z,t) $ describes the amplitude of the electron probability
distribution in the direction transverse to the chain and
can be considered as depending on the continuum variables $y$ and $z$.
Therefore, our Hamiltonian takes the form
\begin{equation}
	\label{HB}
 H \rightarrow H_B =H(P)+
 \frac{P_y^2}{2m_y} +\frac{P_z^2}{2m_z}.
\end{equation}
Here $H(P)$ is the Hamiltonian of the chain (\ref{FH}) where the Peierls transformation
is taken into account (\ref{mom}). In \eqref{HB} $m_x \equiv m$,
$m_y$ and $m_z$ are the corresponding components of the electron effective
mass in the energy band.

Starting from a general gauge for the magnetic field $\vec B=(B_x,B_y,B_z)$
and performing the substitution
$\vec A \rightarrow \vec A + \vec\nabla\Lambda(x,y,z) $ with the proper
choice of the scalar function $\Lambda(x,y,z) $, we can pick the Landau
gauge for the MF, $\vec B=(0,0,B)$, with vector-potential $\vec A=(0,Bna,0)$.
This corresponds to the case where the MF is oriented perpendicular to the
molecular chain and for which, according to the analytical study
\cite{LB-MF-SChFr}, the MF has the most significant impact on
the polaron dynamics. For such
a gauge we get from the Hamiltonian (\ref{HB}) the following system of
equations: 
\begin{eqnarray}
	&&i\hbar \pdt{\Psi_n} +J\left(\Psi_{n+1}+\Psi_{n-1}\right)
	-\bar{\sigma}(u_{n+1}-u_{n-1})\Psi_n
	= \frac{1}{2\,m_y}\left(\hbar k_y-eB\, n\, a\right)^2\Psi_n,\nonumber\\
	&&M \pddt{u_n} = \bar{W}(u_{n+1}-2u_{n}+u_{n-1}) +\bar{\sigma}(|\Psi_{n+1}|^2-|\Psi_{n-1}|^2)
	\label{eq_elphB}
\end{eqnarray}
where the term with $E_0$ (see Eq. \eqref{eq_elphB1}) has been excluded by a phase transformation of
$\Psi_n $ and where $k_y$ is the $y$-component of the electron quasi-momentum.
	
To study this system, it is convenient to perform the following change of
variables to use adimensional parameters:
\begin{eqnarray}
	\tau &=& \frac{Jt}{\hbar}, \quad \xi_n=\frac{u_n}{a},
	\quad \epsilon =\frac{e^2a^2}{2m_yJ},
	\quad \zeta = \frac{2\pi \hbar }{e a},
	\quad \zeta_0 = \frac{\hbar k_y}{e a}=\frac{\zeta}{L_y},\nonumber\\
	\quad \chi &=& \frac{\bar{\sigma} a}{J},
	\quad W = \frac{\hbar^2 \bar{W}}{M J^2},
        \quad \sigma = \frac{\hbar^2 \bar{\sigma}}{M J^2 a}
        \label{eq_rescaling}
\end{eqnarray}
where $L_y$ is the wave length corresponding to the transversal
quasi-momentum $k_y$.
Using \eqref{eq_rescaling}, Eqs. (\ref{eq_elphB}) take the form
\begin{eqnarray}
	&&i \pdtau{\Psi_n}+\left(\Psi_{n+1}+\Psi_{n-1}\right)
	-\chi(\xi_{n+1}-\xi_{n-1})\Psi_n = \epsilon(nB-\zeta_0)^2\Psi_n\nonumber\\
	&&\pddtau{\xi_n} = W(\xi_{n+1}-2\xi_{n}+\xi_{n-1}) + \sigma(|\Psi_{n+1}|^2-|\Psi_{n-1}|^2).
	\label{eq_elphBad}
\end{eqnarray}

Without using perturbative methods, this system of equations does not admit
exact solution. Instead, these equations can be solved
numerically, but to do this we must select some specific system parameters.
From a general point of view, the system parameters not only
affect the solitons properties, they actually decide on the very existence
of the soliton to start with.
Since we study here the impact of the MF on the electron transport
on macroscopic distances, we consider large polarons, {\it i.e.} neither too
small nor almost free electrons.
Therefore, we  study the obtained equations for some set of the
specific numerical values close to those which correspond to
realistic systems of the interest. This is discussed in the next Section.

\section{Choice of system parameters}\label{param}

To select the system parameters we will study, we consider here two classes of
systems, namely electron transport in PP macromolecules and
CPs. It is estimated \cite{Davydov,BPZ-DChA,Gulacsi} that
PPs have the following parameter values: 
\begin{eqnarray}
	&J = (10^{-21}-10^{-20})\,\rm J, \qquad \bar{\sigma} ~= 100 \,{\rm pN}, \quad a= 5.4 \times 10^{-10} \,{\rm m},  \nonumber\\
	&\bar{W} = (39-58) \,{\rm N/m}, \qquad V_{ac} = (3.6-4.5) \times 10^3\,{\rm {m/s}}. 
\end{eqnarray}

It is known that the parameter values of CPs depend on
physical conditions such as temperature, doping and other factors,
but little is known about their numerical values, except, possibly, their
sound velocity (see, e.g.,
\cite{Gulacsi,Le,Aziz} and references therein).  The latter is relatively large 
and varies between 3000 and 9000 m/s in highly dense polymers.
Summarising the available data, we can get some estimates of the
other parameter values.
As a model system we consider here
polypyrrole and  polythiophene. The former has the chemical structure
$\rm{H}\big[[\rm{C}_4\rm{H}2\rm{NH}]_2\big]_n\rm{H}$ and contains two pyrrole
monomers in a unit cell. Similarly, polythiophene, which has the
chemical structure $\rm{H}\big[[\rm{C}_4\rm{H}2\rm{S}]_2\big]_n\rm{H}$,
contains two monomers in a unit cell.
In view of this we consider the following parameter values:
\begin{eqnarray}
		&J =(10^{-21}\,-\,10^{-20}) \, {\rm J}, \quad \bar{\sigma} ~= 100 \, {\rm pN}, \quad a= 8 \times 10^{-10} \,{\rm m}  , \nonumber\\
		\nonumber\\
	&M =216 \times 10^{-27} \,{\rm kg}, \quad
			V_{ac} = (6\,-\,8) \times 10^3\, {\rm m/s} 
\end{eqnarray}	
which gives us the elasticity coefficient
\begin{equation}
	\bar{W}=\frac{MV_{ac}^2}{a^2}=(11.25 \,-\,20.8) \,{\rm N/m}
\end{equation}
and the dimensionless electron-lattice coupling constant
\begin{equation}
	g=\frac{\bar{\sigma}^2}{2J\bar{W}}= (0.24\,-\, 0.77) .
\end{equation}

For this set of the parameter values the polaron is extended over
several lattice sites and the amplitude of its probability in
the absence of external perturbations is in the range 
$ (0.2\,-\,0.4) $.

Aside from studying the effect of MF on a perfectly formed soliton
in an isolated chain, it is also
appropriate to study such effect on a polaron created by a donor complex
on a chain. In these cases, the polarons do not have a perfect bell-like
shape but
can be constituted of a few polarons, propagating at slightly
different velocities. We describe this
system in the next Section.

\section{Donor-Polymer Systems}\label{donor}
\begin{figure}[!ht]
  \centering{
    \includegraphics[width=65mm]{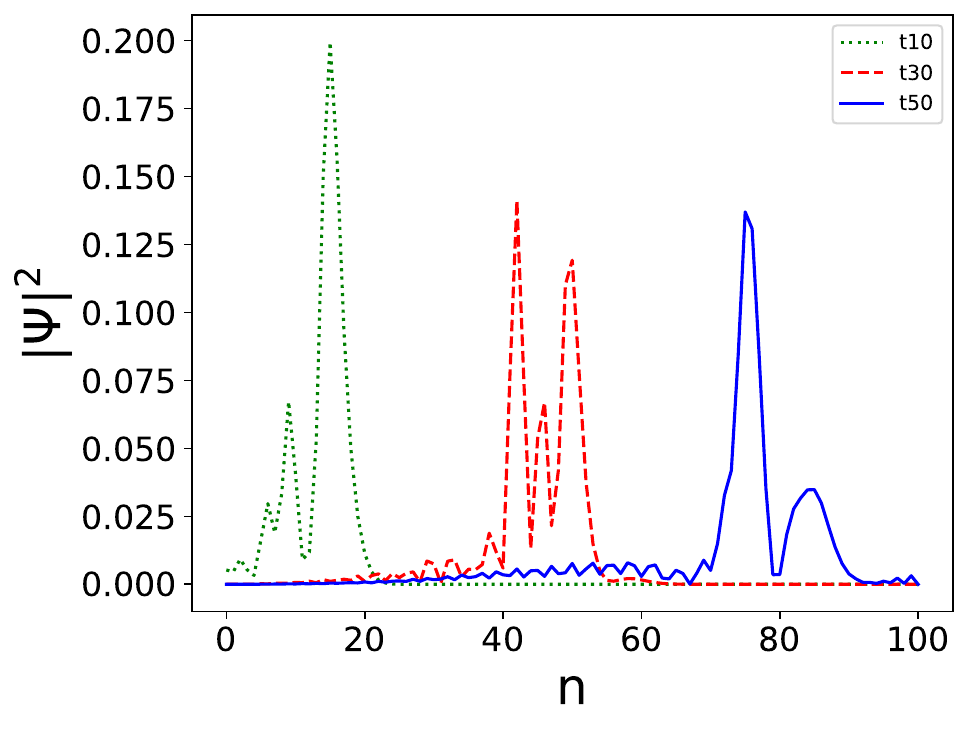}}
  \caption {Time evolution of an Amide-I polaron profile on a Donor-Polypeptide system.
    ($\chi = 174.1$, $W=37$, $\sigma = 0.0857$,
    $D_d=0.6$, $J_d=0.6$, $x_d=0.2$, $v_d=0.6$, $m_d=5$). 
  A small polaron overtakes a larger one.}
\label{fig:AMIDE_profile_mov}      
\end{figure}

After analysing, in Section \ref{num},
the effect of a MF on a polaron created as a static
configuration in the middle of the chain, we will also consider the effect of a
MF on a polaron created by a donor complex \cite{BPZ-DChA}. The
donor, node $0$, is coupled at one end of the chain, node $1$,
so that, after applying the parameter scaling \eqref{eq_rescaling},
the equations are given by

\begin{eqnarray}
i\pdtau{\Psi_{0}}&=&({\cal E}_{0}+D_d)\Psi_{0}-J_{d}\Psi_{1}+
 x_d\chi \left(\xi_1-\xi_0\right)\Psi_{0}+\epsilon\zeta_0^2\Psi_0,
\nonumber\\
i\pdtau{\Psi_{1}}&=&{\cal E}_{0}\Psi_{1}-J^*_{d}\Psi_{0}
- \Psi_{2}
+\chi \left[x_d\left(\xi_1-\xi_0\right) +\left(\xi_2-\xi_1\right)\right] \Psi_{1}
+\epsilon(B-\zeta_0)^2\Psi_1,
\nonumber\\
i\pdtau{\Psi_{n}}&=&{\cal E}_{0}\Psi_{n}
-\left(\Psi_{n+1}+\Psi_{n-1}\right)+\chi \left( \xi_{n+1}-\xi_{n-1} \right)\Psi_{n}+\epsilon(nB-\zeta_0)^2\Psi_n, \quad n =2,\dots, N-1,
\nonumber\\
i\pdtau{\Psi_{N}}&=&{\cal E}_{0}\Psi_{N} -
                     \Psi_{N-1} +\chi \left(\xi_N-\xi_{N-1}\right) \Psi_{N}
                     +\epsilon(NB-\zeta_0)^2\Psi_N.
\label{eq_don_res_psi}
\end{eqnarray}

\begin{eqnarray}
\pddtau{\xi_{0}}&=&\frac{1}{m_d}\left[v_dW(\xi_{1}-\xi_{0}) +x_d\sigma \left(|\Psi_{0}|^2 +|\Psi_{1}|^2 \right)\right],
\nonumber\\
\pddtau{\xi_{1}}&=&v_dW(\xi_{0}-\xi_{1})+W(\xi_{2}-\xi_{1})-x_d\sigma \left(|\Psi_{0}|^2+
|\Psi_{1}|^2\right) +\sigma \left(|\Psi_{1}|^2+|\Psi_{2}|^2 \right),
\nonumber\\
\pddtau{\xi_{n}}&=&W\left( \xi_{n+1}-2\xi_{n}+\xi_{n-1}\right) +
\sigma\left(|\Psi_{n+1}|^2-|\Psi_{n-1}|^2\right), \qquad n =2,\dots, N-1,
\nonumber\\
\pddtau{\xi_{N}}&=&W(\xi_{N-1}-\xi_{N})  -
\sigma \left(|\Psi_{N}|^2 +|\Psi_{N-1}|^2\right),
\label{eq_don_res_u}
\end{eqnarray}
where the index $d$ is used to label the donor parameters and
\begin{eqnarray}
  J_d =&\frac{\bar{J}_d}{\bar{J}}, \,\,\,
  x_d = \frac{\bar{\sigma}_d}{\bar{\sigma}}= \frac{\chi_d}{\chi}, \,\,\,
  v_d = \frac{\bar W_d}{\bar W}, \,\,\,
  m_d = \frac{\bar{M}_d}{\bar{M}}, \,\,\,
\bar{\chi}_d = x_d\chi,\,\,\,
{\cal E}_{0}= \frac{\bar{\cal E}_0}{\bar{J}}. 
\end{eqnarray}
In these Donor-Polymer systems, the electron is initially fully localised on
the donor
molecule, but as time evolves, the electron
generates a few polarons travelling on the chain at slightly different velocities.
This is illustrated in Figure \ref{fig:AMIDE_profile_mov} which shows
the probability density of the polaron at different times, where a small
polaron overtakes a larger one.
The number, size and velocity of these polarons, in general,  depend on
the donor and the chain parameters
as well as the coupling between the chain and the donor.

We have performed simulations for a number of donor parameters:
$D_d$ was scanned for values in the range $0.1$ to $2$ by step of
$0.1$. $J_d$, $X_d$ and $V_d$ were all scanned for values in the
range $0.1$ to $1$ by step of $0.1$. $M_d$ was scanned for values in the
range $1$ to $5$ by step of $0.5$. For each type of polaron, we identified
the donor parameters
leading to the most localised polaron reaching the end of the lattice. 
The values we obtained are given in table \ref{table1}.

We then ran simulations with
various values of the MF $B$ and looked at the profile of the
polaron when it reaches the end of the lattice chain
which we chose to contain 100 nodes.  
We also set the transverse quasi-momentum $k_y$ to a range of values.

\begin{table}[ht!]
\centerline{
  \begin{tabular}{llllll} 
$x_d$&$x_a$ &$D_d$ & $J_d$& max $|\Psi|^2$& $N$\\ 
\hline
0.1&0.1 &0.6 &0.7 &0.183&0.75\\
0.3&0.3 &0.6 &0.7 &0.201&0.81\\
0.6&0.6 &0.8 &0.7 &0.219&0.82\\
1 &1 &2.2 &0.9 &0.251&0.89\\
\end{tabular}}
\caption{Optimal parameter values for the formation of a polaron 
on the chain with
$v_d=v_a=0.1$, $W=0.88$, $\chi=0.8$ and $m_d=3$.}
\label{table1} 
\end{table}	
	
\section{Results of numerical simulations}\label{num}
To solve Eqs. \eqref{eq_elphBad} numerically we first rewrote the equation
for $\xi_n$ as a system of pair of first order differential equations
\begin{eqnarray}
\pdtau{\xi_n} &=& \Xi_n\nonumber\\
\pdtau{\Xi_n}&=& W(\xi_{n+1}-2\xi_{n}+\xi_{n-1}) +\sigma(|\Psi_{n+1}|^2-|\Psi_{n-1}|^2).\nonumber\\
	\label{eq_elphBad_num}
\end{eqnarray}

The numerical simulations were performed by first computing the profile for
the polaron by solving Eqs. \eqref{eq_elphBad} to which an absorbing term has been
added and with $B=\zeta_0=0$. This was then saved as an initial profile used
for further simulations, solving Eqs. \eqref{eq_elphBad} with non zero $B$ and
$\zeta_0$. 

At the start of some simulations we also boosted the electron to velocity $V$
by applying the following transformation on $\Psi_n$ and $\Xi_{n}$ 
\begin{eqnarray}
  \Psi_{n}^{(B)} &=& \Psi_n\exp(i\,V\,(n-n_0))\nonumber\\
  \Xi_{n}^{(B)} &=& (\xi_{n-1}-\xi_{n+1})\sin(V).
\end{eqnarray}

For the numerical simulations, we consider three different specific systems:
an Amide-I vibration polaron in a PP chain, an extra electron in a
PP and an electron in a CP. The parameter values
used are listed in Table \ref{tab_param1} while the corresponding
adimensional parameters are given in Table \ref{tab_adim_param1}.

We now describe our results for the three cases considered.

\begin{table}[htb!]
\begin{tabular}{|l|l|l|l|}
  \hline
  Parameter & Amide-I & Extra Electron & Conducting Polymer \\
  \hline
  $J$ (J) & $1.55\times 10^{-22}$& $ 1.55\times 10^{-22}$ & $1\times 10^{-21} $\\
 $\bar{\sigma}$(N) & $50\times 10^{-12}$& $50\times 10^{-12} $ & $100\times 10^{-12} $ \\
$\bar{W}$(N/m) & $40$& $40 $ & $28 $ \\
$M$(kg) & $500\times 10^{-27}$& $ 500\times 10^{-27}$ & $500\times 10^{-27} $ \\
$a$(m) & $5.4\times 10^{-10} $& $5.4\times 10^{-10}  $ & $5.4\times 10^{-10}  $ \\
  \hline
\end{tabular}
\caption{Dimensional parameters for the three systems considered}
\label{tab_param1}
\end{table}

\begin{table}[htb!]
\begin{tabular}{|l|l|l|l|}
  \hline
  Parameter & Amide-I & Extra Electron & Conducting Polymer \\
  \hline
$\chi $& $174.19$ & $54 $ & $80 $\\
$\epsilon$ & $1.963\times 10^{-7}$& $ 1.963\times 10^{-7}$ & $ 9.4542\times 10^{-7}$\\
$\zeta ({\rm Tm} )$& $7.6586\times 10^{-6}$& $7.6586\times 10^{-6} $ & $5.1696\times 10^{-6} $\\
$W$        & $37$& $0.62 $ & $0.628 $\\
$\sigma$ & $0.0857$& $0.0041 $ & $0.00646 $\\
  \hline
\end{tabular}
\caption{Adimensional parameters for the thee systems considered.
  The time scale $\tau$ corresponds to $0.68$ps for the Amide-I polaron and
$0.1$ps for the extra electron in a polypeptide or conducting polymer.}
\label{tab_adim_param1}
\end{table}

\subsection{Amide-I type polaron in polypeptides}
\begin{figure}[!ht]
  \centering{
    \includegraphics[width=60mm]{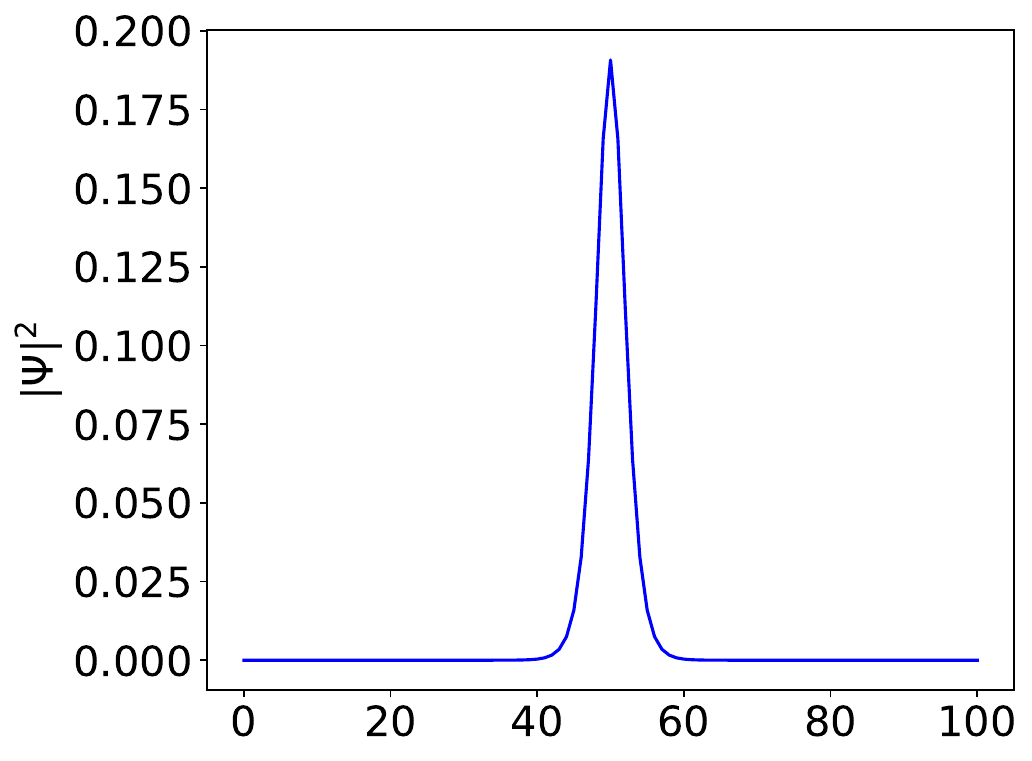}\quad
    \includegraphics[width=60mm]{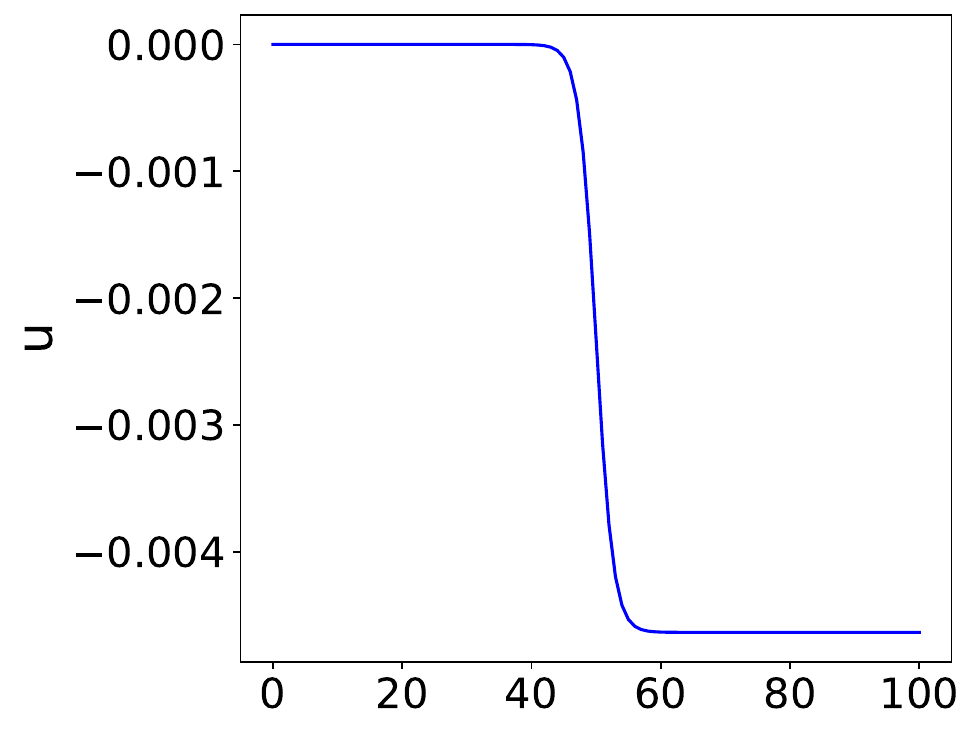}}\\
  \centering{ a) \hspace{60mm} b)}
  \caption {Amide-I type polaron ($\chi = 174.19$, $W=37$, $\sigma = 0.0857$)
    as a function of $n$ : a) profile $|\Psi|^2$; b) site displacement $u$.}
\label{fig:AMIDE_profiles}      
\end{figure}

The Amide-I type polaron is localised with a half width of approximately 4
lattice sites as shown in Figure \ref{fig:AMIDE_profiles}.

\subsubsection{The role of polaron boosts}
\begin{figure}[!ht]
  \centering{
    \includegraphics[width=80mm]{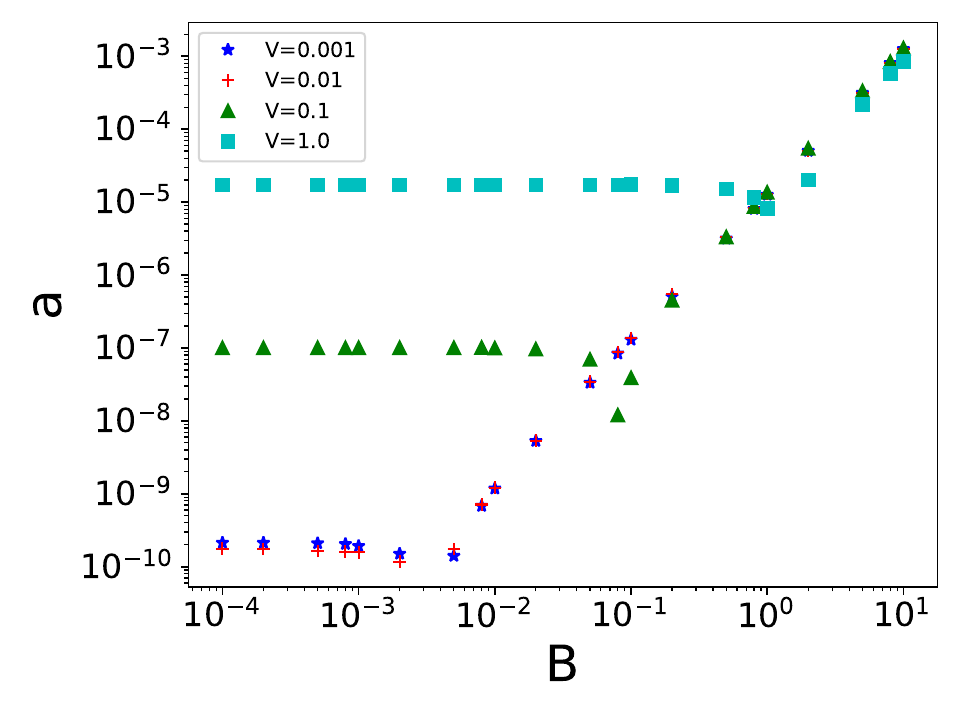}}
    \caption {Amide-I type polaron: acceleration as a function of $B$ for different boosting velocities.}
\label{fig:AMIDE_a_boost}      
\end{figure}

The critical velocity, above which the polaron starts accelerating,
is approximately $4.4\times  10^{-5}$ in dimensionless
units. The acceleration depends on the initial boost velocity and the MF.
Below the critical velocity, the polaron starts to move when
$B\ge 0.2 {\rm T}$. Above the critical velocity the acceleration varies
approximately from $5\times 10^{-7}$ for $B=0.2{\rm T}$ to $1.2\times 10^{-3}$
for $B=10{\rm T}$ and to larger value when the boosting velocity is above the
critical value, as shown in Figure \ref{fig:AMIDE_a_boost}.

\subsubsection{Impact of the transverse quasi-momentum}
\begin{figure}[!ht]
  \centering{
    \includegraphics[width=65mm]{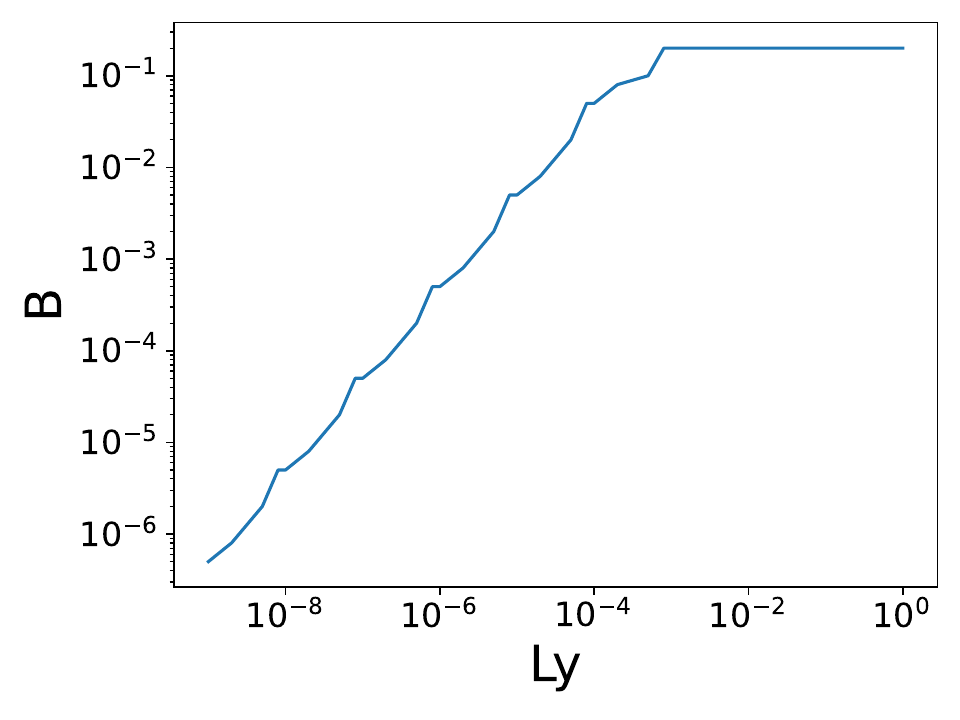}
    \includegraphics[width=65mm]{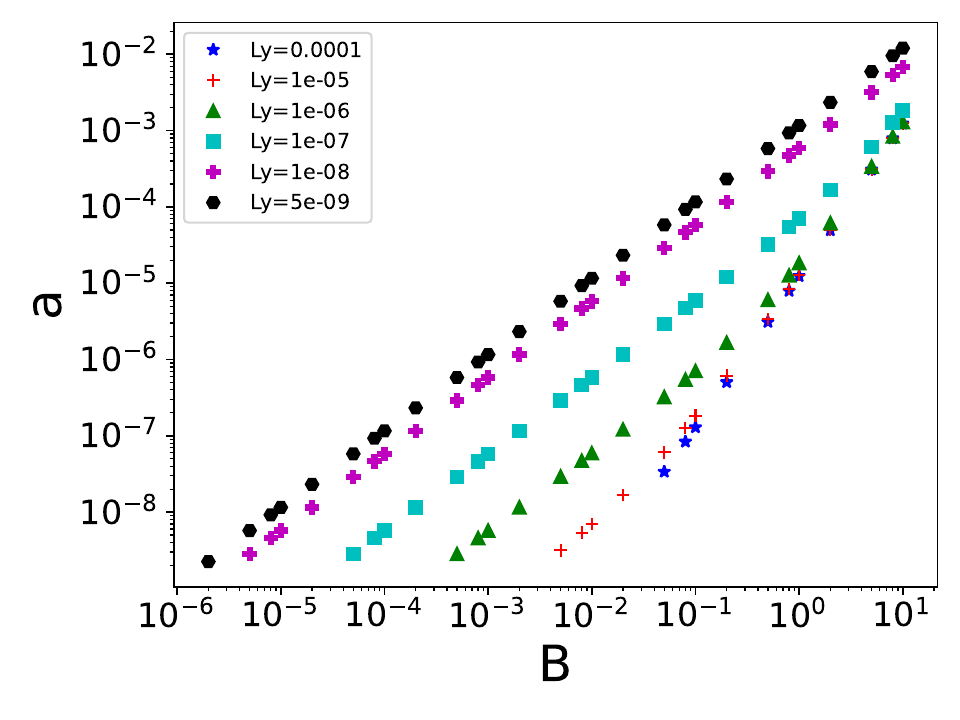}}\\
  \centering{a) \hspace{65mm} b)}
  \caption {Amide-I polaron ( $\chi = 174.19$, $W=37$, $\sigma = 0.0857$)
    a) : critical $B$ as a function of $L_y$;
  b) acceleration as a function of $B$ for different $L_y$.}
\label{fig:AMIDE_crit_B_a_Ly}      
\end{figure}

At non zero transverse quasi-momentum,
the critical MF required
to make the polaron move depends on the transverse quasi-momentum, or
correspondingly on the wavelength $L_y$.
The critical MF is a linear function of $L_y$ when
$L_y< 8\times 10^{-4}{\rm m}$.
When $L_y\ge 8\times 10^{-4}{\rm m}$, the polaron only moves if $B\ge 0.2{\rm T}$. 
For smaller values of $L_y$ is starts moving for smaller values of $B$,
as small as $B=5 \times 10^{-7}{\rm T}$ when $L_y=10^{-9}{\rm m}$. 
This is illustrated in Figure
\ref{fig:AMIDE_crit_B_a_Ly}.a. The little kinks in the graph are artefacts of the
sampling of $B$ and $L_y$ values
which we have used. The acceleration depends on $L_y$ and it is a nearly
linear function of $B$ (Figure \ref{fig:AMIDE_crit_B_a_Ly}.b).

\subsection{Extra electron in polypeptides}
The electron polaron in a PP  is delocalised with a half width of
approximately 5 lattice sites. The polaron profile is very similar to the one shown
in Figure \ref{fig:AMIDE_profiles} except that the maximum of $|\Psi|^2$ is $0.17$.

\subsubsection{Boost}
\begin{figure}[!ht]
  \centering{
    \includegraphics[width=80mm]{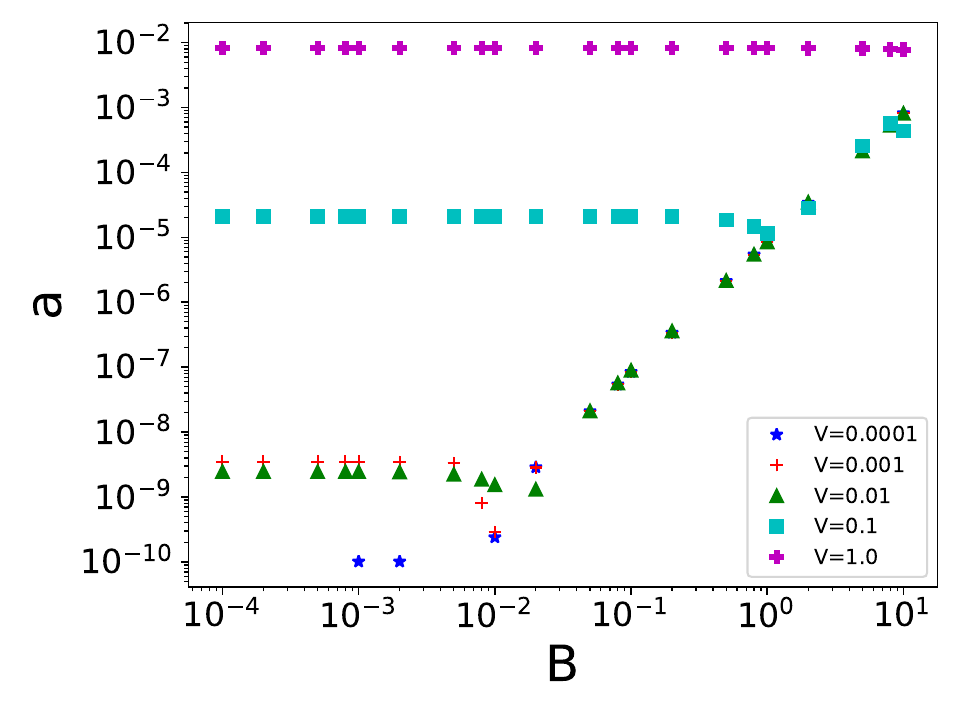}}
  \caption {Extra electron in a polypeptide $(\chi = 54$, $W=0.62$, $\sigma = 0.0041$):
    acceleration as a function of $B$ for different boosts}
\label{fig:AMIDE_EE_a_boost}      
\end{figure}

The critical boost velocity is approximately $2.4\times  10^{-5}$ in dimensionless
units.
The acceleration depends on the initial boost velocity and the MF.
Below the critical velocity, the polaron starts to move when $B\ge 0.05 {\rm T}$.
Below the critical velocity the acceleration varies approximately from
$2\times 10^{-8}$
for $B=0.05{\rm T}$ to $8 \times 10^{-4}$ for $B=10{\rm T}$ and to larger value
when the
boost velocity is above the critical value, as shown in Figure
\ref{fig:AMIDE_EE_a_boost}.

\subsubsection{Impact of the polaron transverse quasi-momentum}
\begin{figure}[!ht]
  \centering{
    \includegraphics[width=65mm]{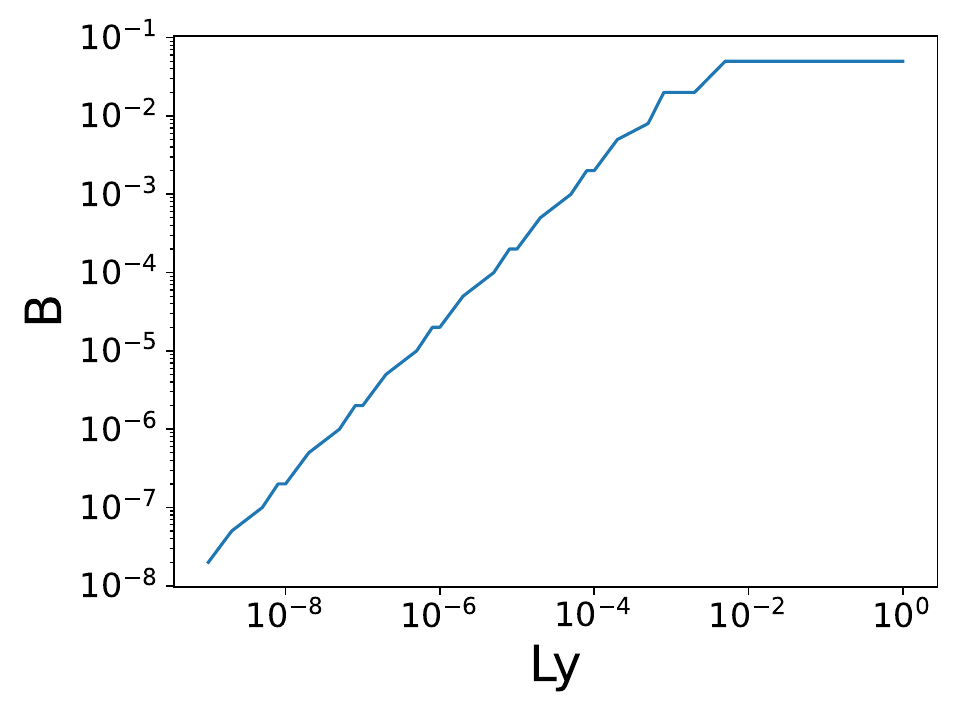}
    \includegraphics[width=65mm]{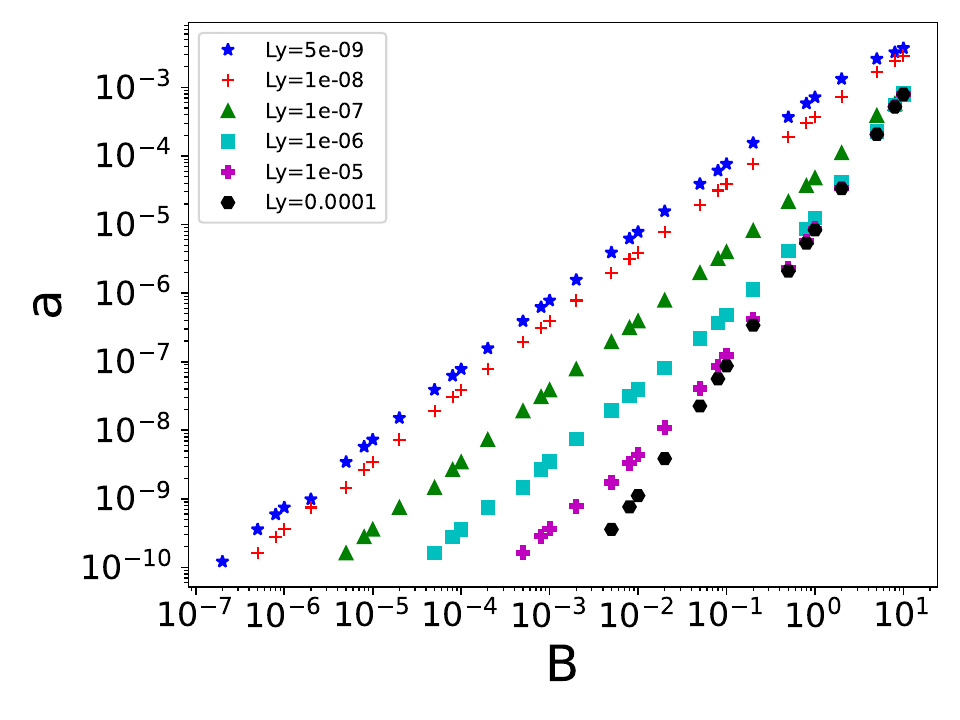}}\\
  \centering{a) \hspace{65mm} b) }
  \caption {Extra electron in a polypeptide ($\chi = 54$, $W=0.62$, $\sigma = 0.0041$):
    a) critical $B$;
    b) acceleration as a function of $B$ for different $L_y$.}
\label{fig:AMIDE_EE_crit_B_a_Ly}      
\end{figure}

At non zero transverse quasi-momentum, the critical MF required
to make the polaron move depends on the wavelength $L_y$ corresponding
to the transverse quasi-momentum. The critical MF is a linear function of $L_y$ when
$L_y< 0.002{\rm m}$.
When $L_y\ge 0.002 {\rm m}$, the polaron only moves if $B\ge 0.05{\rm T}$. 
For smaller values of $L_y$ is starts moving for smaller values of $B$,
as small as $B=2 \times 10^{-8}{\rm T}$ for $L_y= 10^{-9}{\rm m}$.

This is illustrated in Figure \ref{fig:AMIDE_EE_crit_B_a_Ly}.a.
The acceleration depends on $L_y$ and it is a nearly linear function of $B$
(Figure \ref{fig:AMIDE_EE_crit_B_a_Ly}.b).

\subsection{Conducting Polymer}
The CP polaron is delocalised with a half width of
approximately 3 lattice sites.
The polaron profile is very similar to the one shown
in Figure \ref{fig:AMIDE_profiles} except that the maximum of $|\Psi|^2$ is $0.34$.

\subsubsection{The role of polaron boosts}
\begin{figure}[!ht]
  \centering{
    \includegraphics[width=80mm]{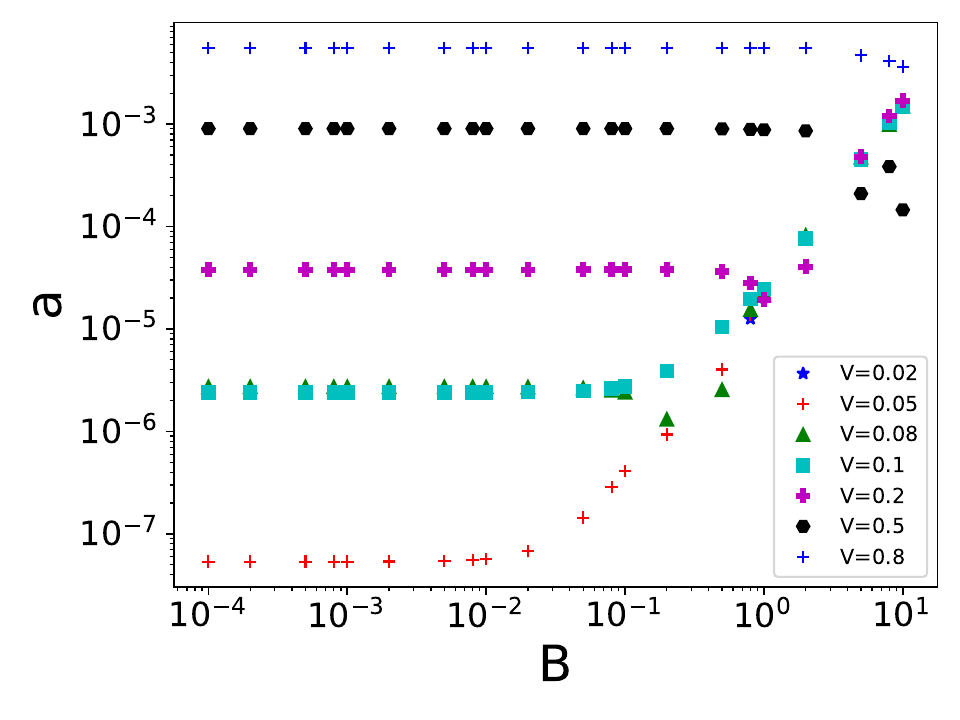}}
  \caption {Conducting polymer: absolute value of the acceleration as a
    function of $B$ for different boosts}
\label{fig:CP_a_boost}      
\end{figure}

The critical boost velocity is approximately $0.021$ in dimensionless units.
As shown in Figure \ref{fig:CP_a_boost},
below the critical velocity, the polaron never accelerates even when
$B=10{\rm T}$,
except when the boosting velocity is slightly below the critical value
($V=0.02$).
Above the critical value, the polaron starts to accelerate when
$B\le 0.8 {\rm T}$.
The acceleration varies then approximately from $10^{-5}$ for $B=0.8{\rm T}$ to
$1.5\times 10^{-3}$ for $B=10{\rm T}$ as shown in
Figure \ref{fig:CP_a_boost}.
Surprisingly the polaron moves in the direction opposite to the
boosting direction except when the boosting velocity is slightly
below the critical
velocity and $B$ is close to the minimum value to make the polaron move. 

\subsubsection{Impact of the transverse quasi-momentum}

\begin{figure}[!ht]
  \centering{
    \includegraphics[width=65mm]{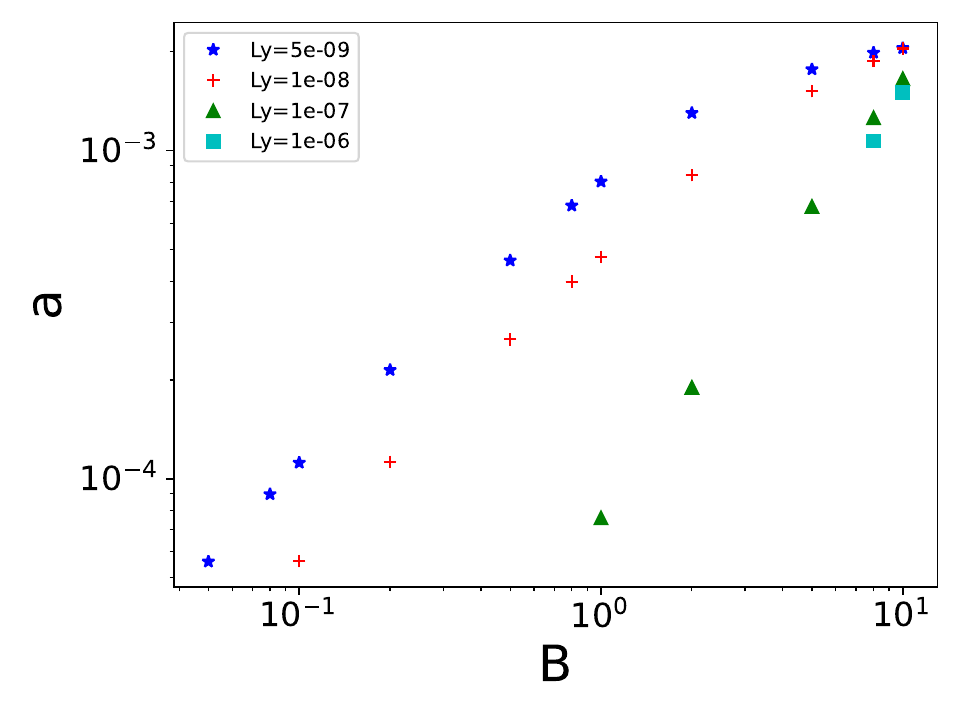}
    \includegraphics[width=65mm]{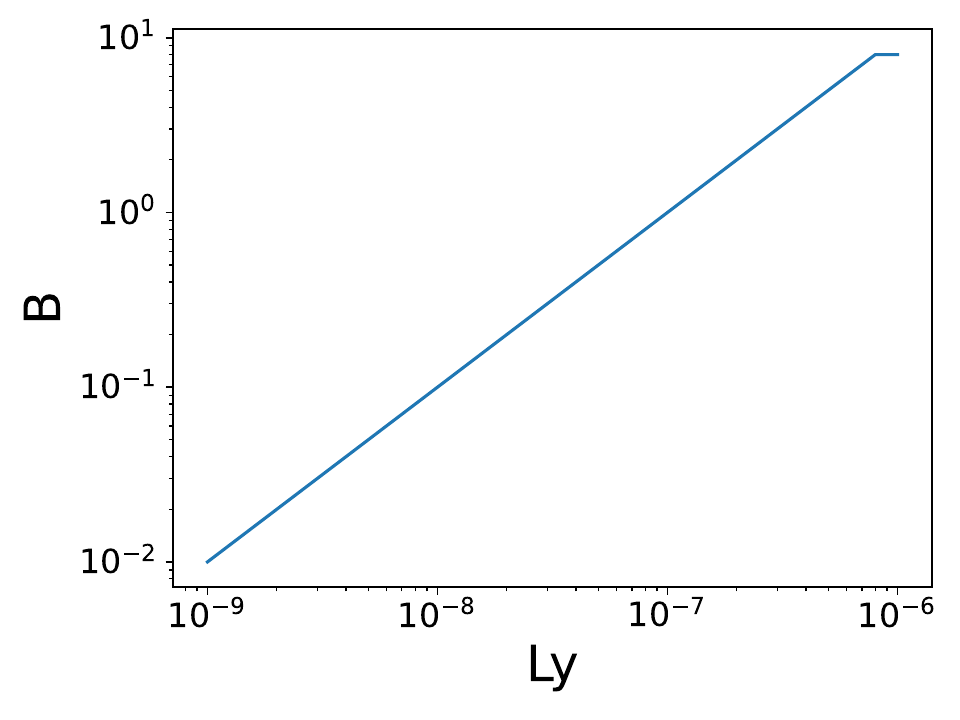}}\\
  \centering{a) \hspace{65mm} b)}
  \caption {Conducting polymer ($\chi = 80$, $W=0.628$, $\sigma = 0.00646$) :
    a) acceleration as a function of $B$ for different $L_y$;
    b) critical $B$ as a function of $L_y$}
\label{fig:CP_crit_B_a_Ly}      
\end{figure}

At non zero transverse quasi-momentum, the critical MF required
to make the polaron move depends on the wavelength $L_y$ and it is a linear
function of $L_y$. When $L_y\ge2\times 10^{-6}{\rm m}$
one needs a MF larger than $10{\rm T}$ which we haven't tested.
When $L_y= 5\times 10^{-9}{\rm m}$, one needs a MF of at least $0.05{\rm T}$
to move the polaron. This is illustrated in Figure
\ref{fig:CP_crit_B_a_Ly}.a. 
The acceleration depends on $L_y$ and it is a nearly linear function of $B$
(Figure \ref{fig:CP_crit_B_a_Ly}.b).

\subsection{Amide-I polaron in a Donor-Polypeptide system}
For Amide-I polarons in Donor-Polypeptide systems we have chosen the
following parameters:
\begin{eqnarray}
&&\chi = 174.10,\,\,\,
\epsilon = 1.963\times 10^{-27},\,\,\,
\zeta= 7.6586\times 10^{-6}\,m{\rm T},\,\,\,
W = 37,\nonumber\\
&&\sigma = 0.0857,\,\,\,
D_d=0.6,\,\,\, J_d=0.6,\,\,\, x_d=0.2,\,\,\, v_d=0.6,\,\,\, m_d=5.  
\end{eqnarray}

\begin{figure}[!ht]
  \centering{
    \includegraphics[width=65mm]{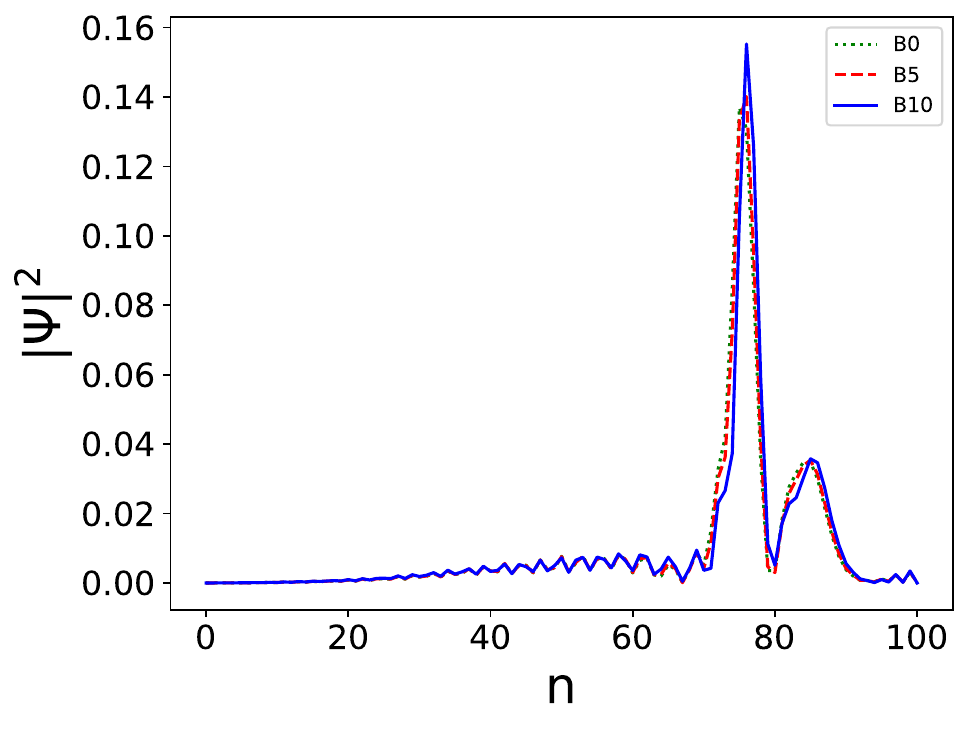}
    \includegraphics[width=65mm]{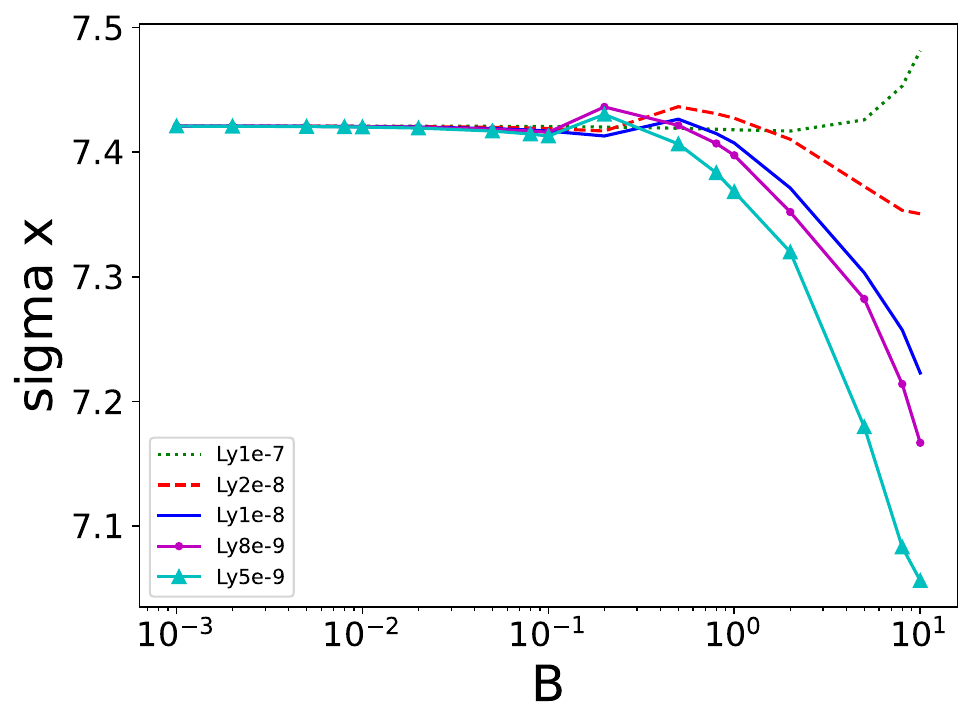}
  }
  \centering{a) \hspace{65mm} b)}
  \caption {Amide-I polaron 
    as a function of $B$  ($\chi = 174.1$, $W=37$, $\sigma = 0.0857$,
  $D_d=0.6$, $J_d=0.6$, $x_d=0.2$, $v_d=0.6$, $m_d=5$):
a) Polaron profile after 50 units of time;
b) Width, standard deviation, of the polaron after 30.8 units of time for different values of $L_y$.}
\label{fig:AMIDE_profiles_sigma_B}      
\end{figure}

Figure \ref{fig:AMIDE_profiles_sigma_B}.a shows several profiles of the polaron after
50 units of time when the transverse quasi-momentum is negligibly small,
$L_y >>$. We see that
the effect of the MF on the polaron is negligible even when $B=10{\rm T}$. 
Figure \ref{fig:AMIDE_profiles_sigma_B}.b
shows the average width of the polaron after reaching the middle of the lattice. We see
that the polaron width does not change significantly in relatively weak MF
and that
for larger MF it depends on the transverse quasi-momentum.

\begin{figure}[!ht]
  \centering{
    \includegraphics[width=60mm]{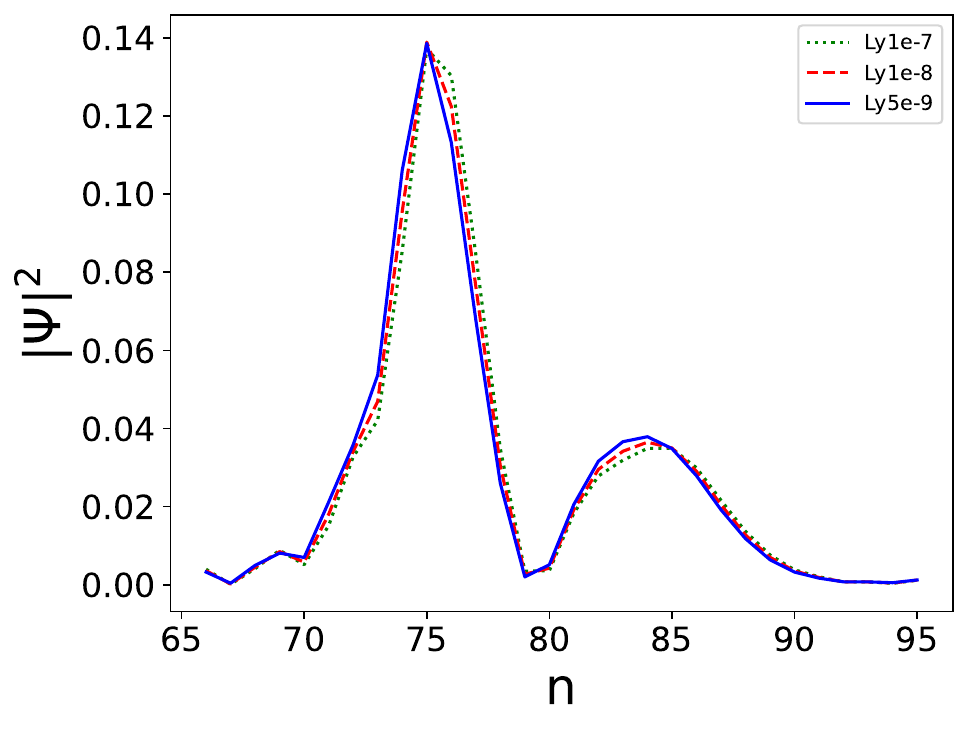}
    \includegraphics[width=60mm]{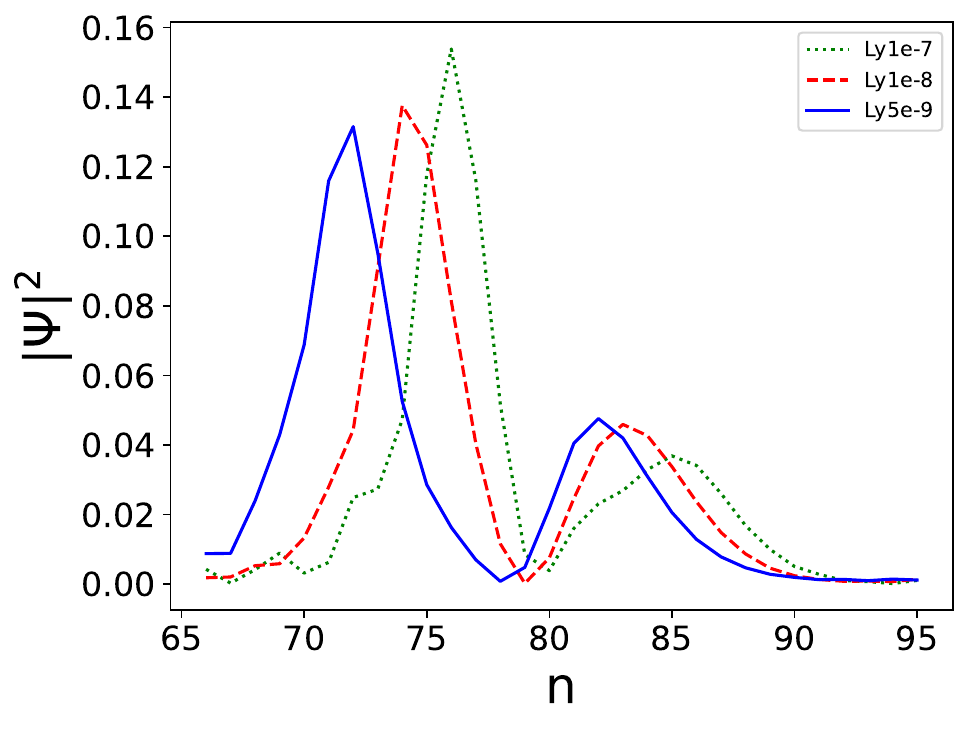}
  }
  
  \centering{a)\hspace{60mm}b)}

  \centering{
    \includegraphics[width=60mm]{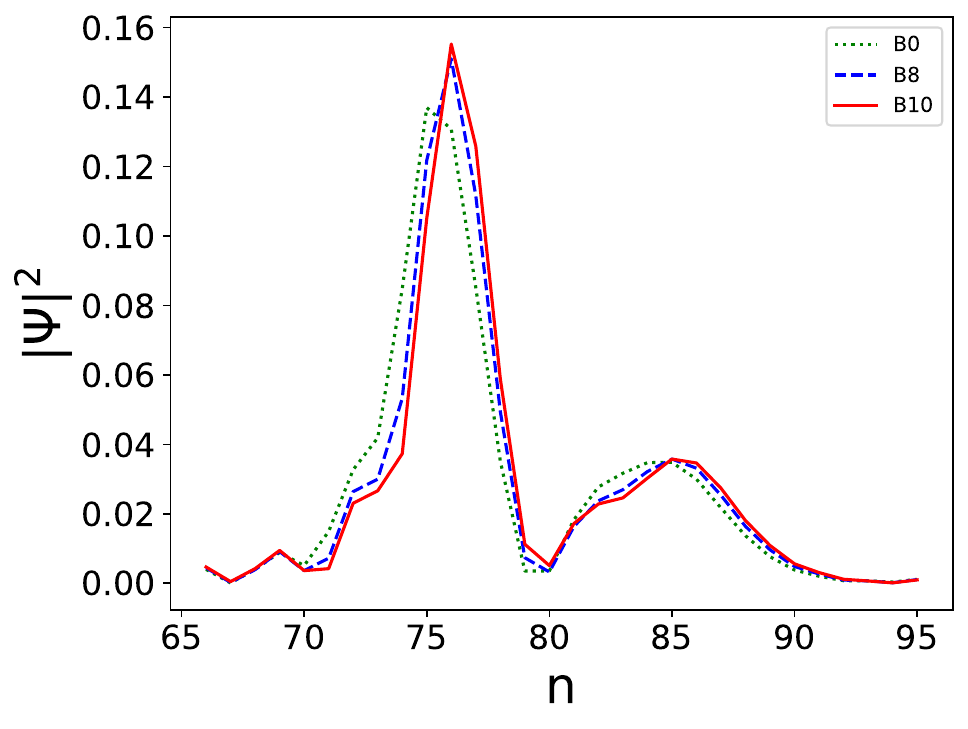}
    \includegraphics[width=60mm]{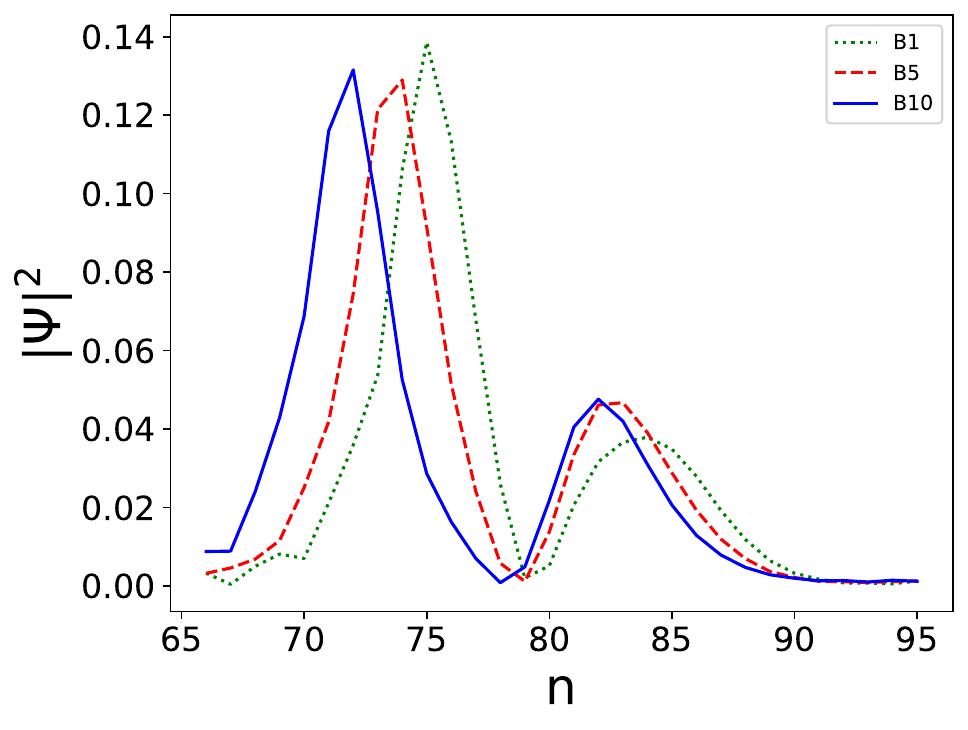}
  }
  
  \centering{c)\hspace{60mm}d)}
  \caption {Profile of the Amide-I polaron in a Donor-Polypeptide system
    after 50 units of time as functions of $B$ and $L_y$.
    $\chi = 174.1$, $W=37$, $\sigma = 0.0857$,
  $D_d=0.6$, $J_d=0.6$, $x_d=0.2$, $v_d=0.6$, $m_d=5$:
  a) $B=1$; b) $B=10$; c) $k_y=0$; d) $L_y=5\times 10^{-9}$ (Restricted to the end of the lattice).
}
\label{fig:Amide-1_profiles_B_Ly}      
\end{figure}
As we can see from Figure \ref{fig:Amide-1_profiles_B_Ly}, weak MFs
only have a little impact on the polaron profile even when $L_y$
is very small. This impact increases with increasing $B$
(Figure \ref{fig:Amide-1_profiles_B_Ly} ). The dependence of the
soliton width on the transverse quasi-momentuam also increases with $B$.
When $B$ is large, {\it i.e.} well above 1T, the polaron
is slightly slowed down (Figure \ref{fig:Amide-1_profiles_B_Ly}.d).

\subsection{Electron in a Donor-Polypeptides system}
For the electron polaron in a Donor-Popylpeptide system we have chosen
the following parameters:
\begin{eqnarray}
&&\chi = 54,\,\,\,
\epsilon = 1.963\times 10^{-7},\,\,\,
\zeta = 7.6586\times 10^{-6}\, m{\rm T}\,\,\,
W = 0.62\nonumber\\
&&\sigma = 0.0041,\,\,\,
D_d=0.1,\,\,\, J_d=0.6,\,\,\, x_d=0.1,\,\,\, v_d=0.2,\,\,\, m_d=5.  
\end{eqnarray}

When the transverse quasi-momentum is very small, ($L_y >>$),
the effect of the MF is negligible
(not shown) even when $B=10{\rm T}$.

\begin{figure}[!ht]
  \centering{
    \includegraphics[width=100mm]{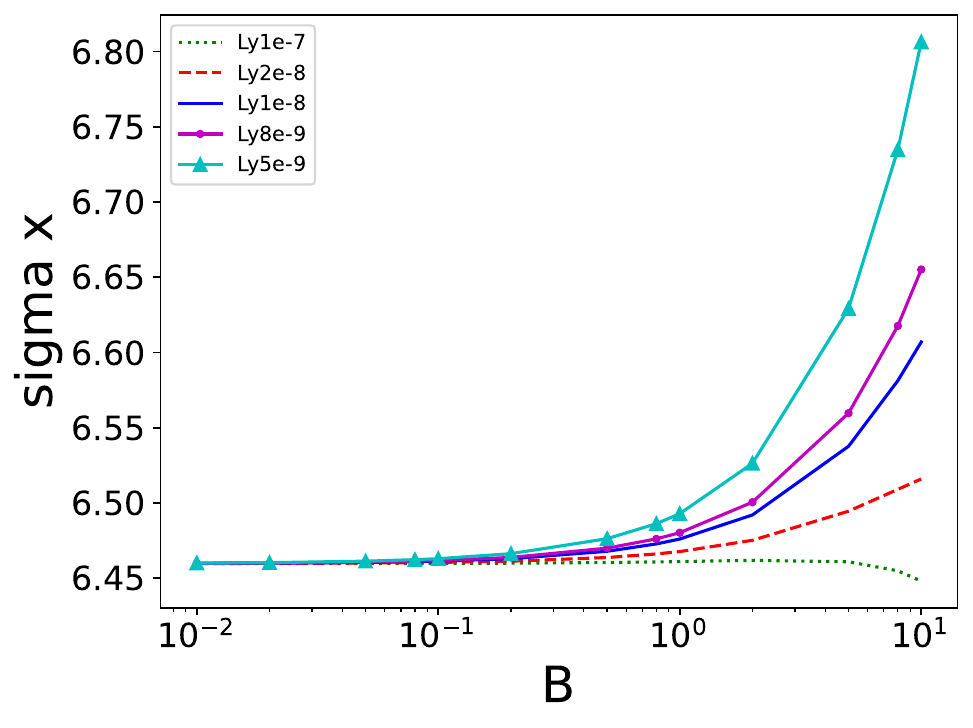}
}
\caption {Width, standard deviation, of the polaron in the Donor-Polypeptide
  after 27.45 units of time
   ($\chi = 54$, $W=0.62$, $\sigma = 0.0041$,
    $D_d=0.1$, $J_d=0.6$, $x_d=0.1$, $v_d=0.2$, $m_d=5$).}
\label{fig:AMIDE_EXE_sigmax_B}      
\end{figure}

Figure \ref{fig:AMIDE_EXE_sigmax_B}
shows the average width of the polaron profile after 27.45 units of time and we
see that as $B$ increases, the polaron becomes broader and this change of
polaron width increases as the transverse quasi-momentum increases,
but remains relatively small.

\begin{figure}[!ht]
  \centering{
    \includegraphics[width=60mm]{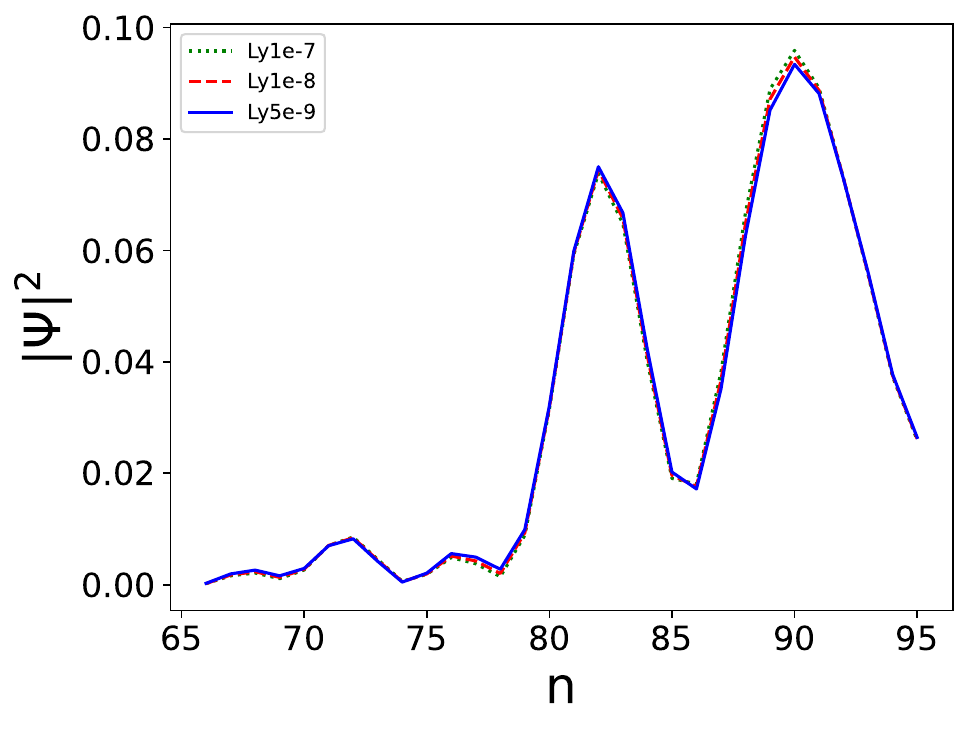}
    \includegraphics[width=60mm]{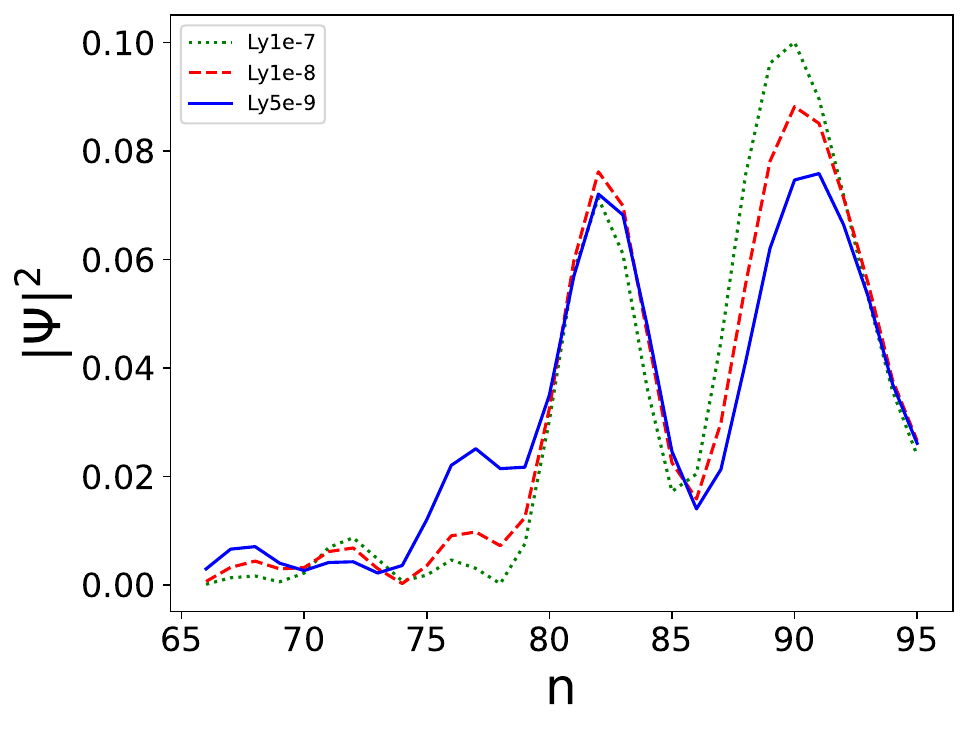}
  }
  
  \centering{a)\hspace{60mm}b)}

  \centering{
    \includegraphics[width=60mm]{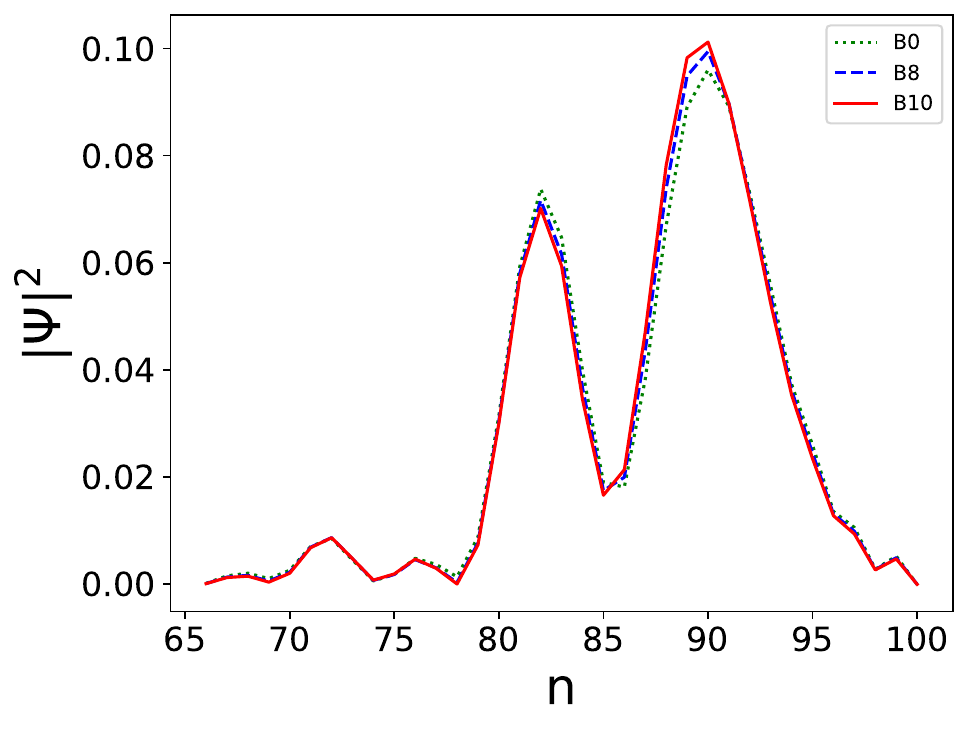}
    \includegraphics[width=60mm]{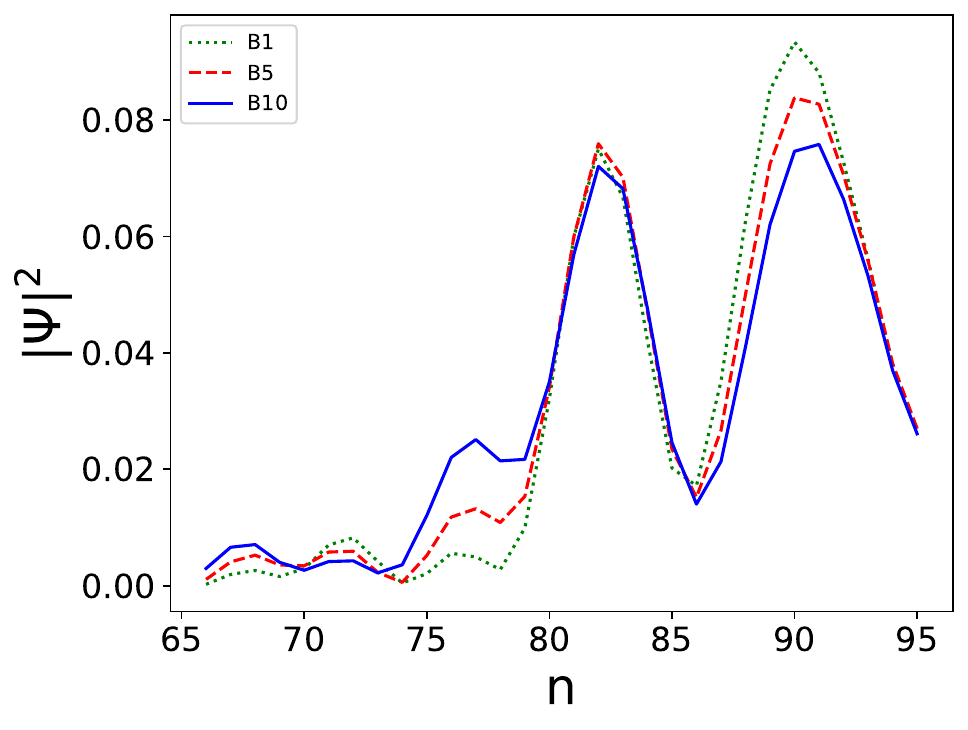}
  }
  
  \centering{c)\hspace{60mm}d)}
  \caption {Profile of the polaron in a Donor-Polypeptide system
    after 48 units of time for several values of the magnetic field $B$.
  $\chi = 54$, $W=0.62$, $\sigma = 0.0041$,
    $D_d=0.1$, $J_d=0.6$, $x_d=0.1$, $v_d=0.2$, $m_d=5$.  
  a) $B=1$; b) $B=10$; c) $k_y=0$; d) $L_y=5\times 10^{-9}$. 
}
\label{fig:AMIDE_EXE_profiles_B_Ly}      
\end{figure}

As we can see from Figure \ref{fig:AMIDE_EXE_profiles_B_Ly}, the MF
has a very negligible effect on the polaron profile even when $L_y$
is very small. When $B$ is very large, {\it i.e.} well above 1T, the
polaron is slightly slowed down (Figure \ref{fig:AMIDE_EXE_profiles_B_Ly}.d).

\subsection{Donor-Conducting Polymer System}
\begin{eqnarray}
&&\chi = 80,\,\,\,
\epsilon = 9.4542\times 10^{-7},\,\,\,
\zeta = 5.1696\times 10^{-6}\,m{\rm T}\,\,\,
W = 0.628,\nonumber\\
&&\sigma = 0.00646,\,\,\,
D_d=0.1,\,\,\, J_d=0.5,\,\,\, x_d=0.3,\,\,\, v_d=0.1,\,\,\, m_d=5.  
\end{eqnarray}

\begin{figure}[!ht]
  \centering{
    \includegraphics[width=65mm]{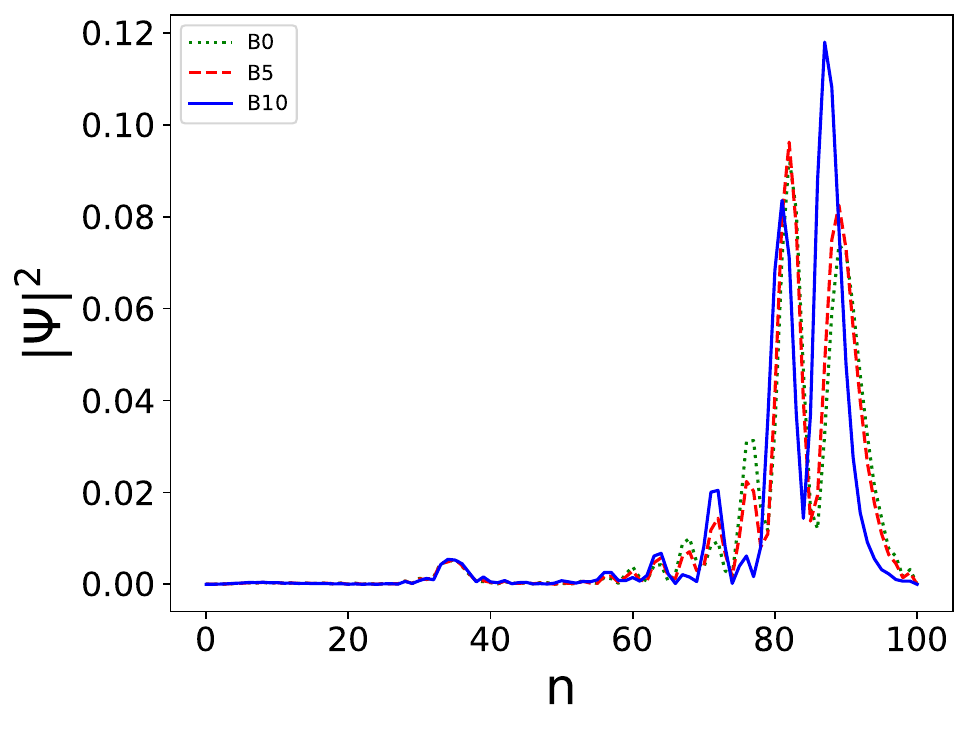}
    \includegraphics[width=65mm]{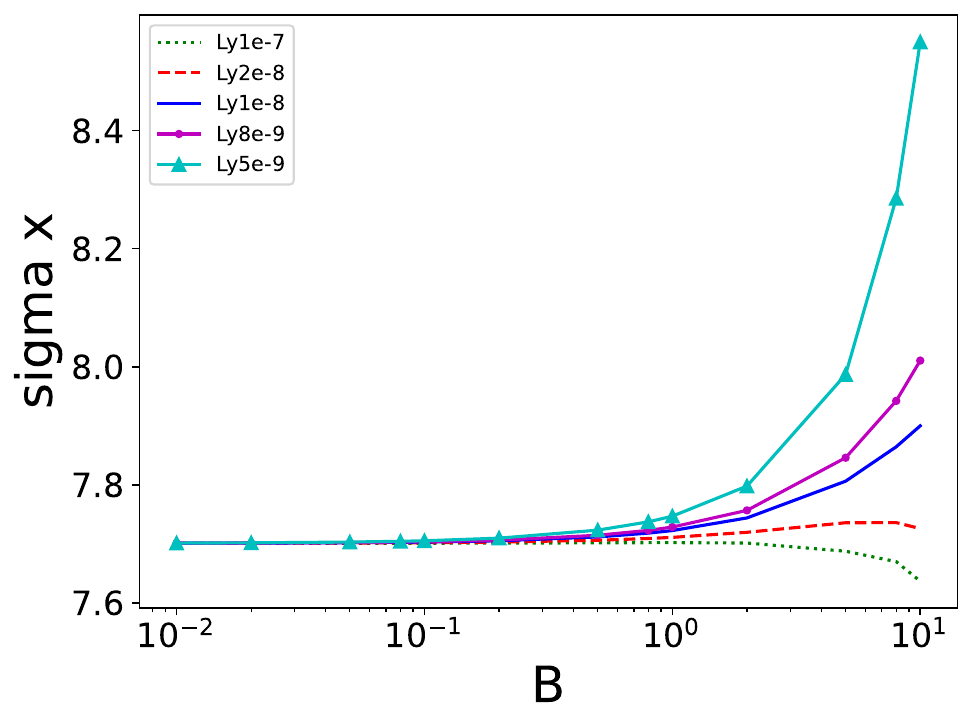}
  }
  \centering{a)\hspace{65mm}  b)}
  \caption {Polaron in a Donor-Conducting Polymer system 
    as a function of $B$ when $k_y=0$
    ($\chi = 80$, $W=0.628$, $\sigma = 0.00646$,
  $D_d=0.1$, $J_d=0.5$, $x_d=0.3$, $v_d=0.1$, $m_d=5$):
  a) Polaron profile after 48 units of time;
  b) Width, standard deviation, of the polaron after 28 units of time for different values of $L_y$.
}
\label{fig:CP_profiles_sigmax_B}      
\end{figure}

Figure \ref{fig:CP_profiles_sigmax_B}.a shows several profiles of the polaron after 48
units of time when the transverse quasi-momentum is very small. We see
that the effect
of the MF on the polaron profile is negligible even when $B=10{\rm T}$. 
Figure \ref{fig:CP_profiles_sigmax_B}.b
shows that the average width of the polaron profile after
reaching the middle of the lattice
does not vary much as a function of  $B$ in weak MFs but increases
in strong MFs.
The broadening of the polaron
increases as the transverse quasi-momentum becomes larger but the effect
remains relatively small.

\begin{figure}[!ht]
  \centering{
    \includegraphics[width=60mm]{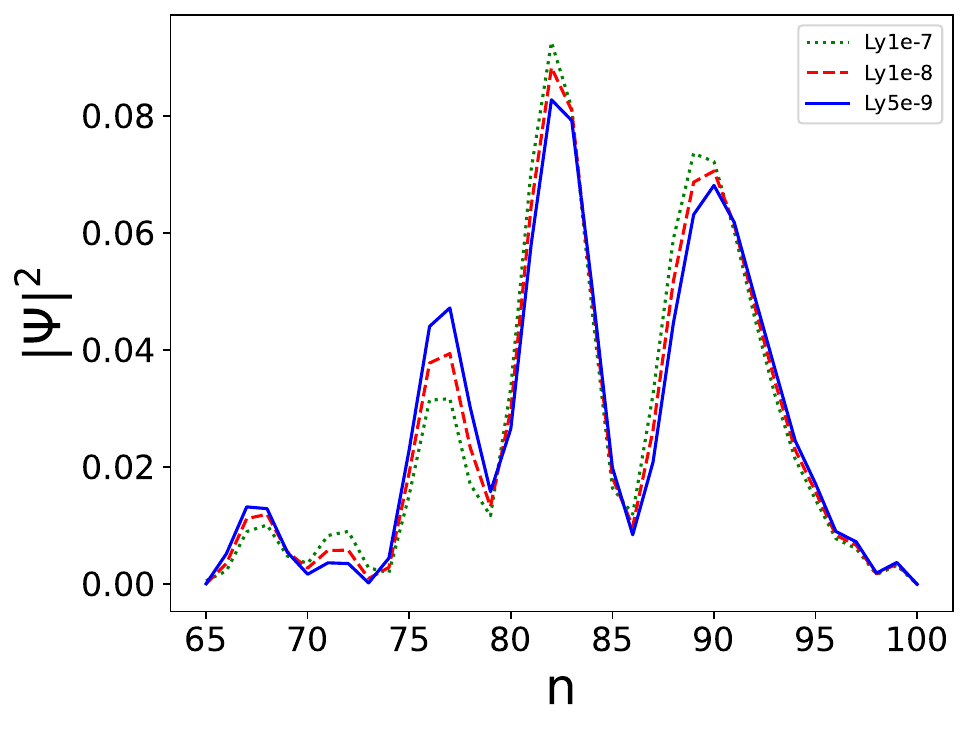}
    \includegraphics[width=60mm]{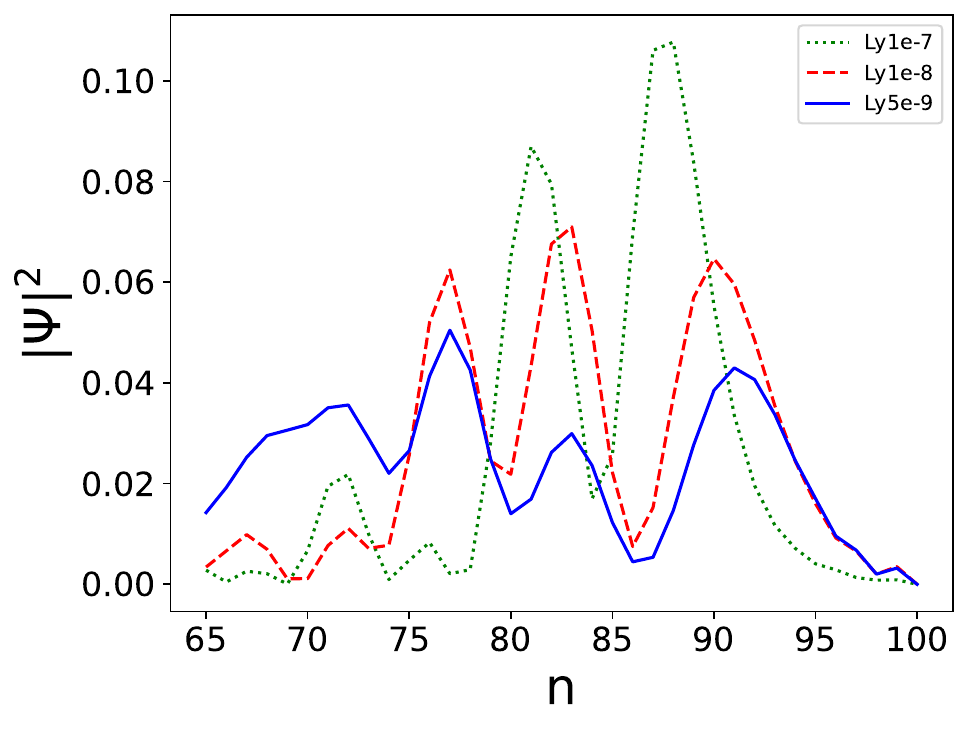}
  }
  
  \centering{a)\hspace{60mm}b)}

  \centering{
    \includegraphics[width=60mm]{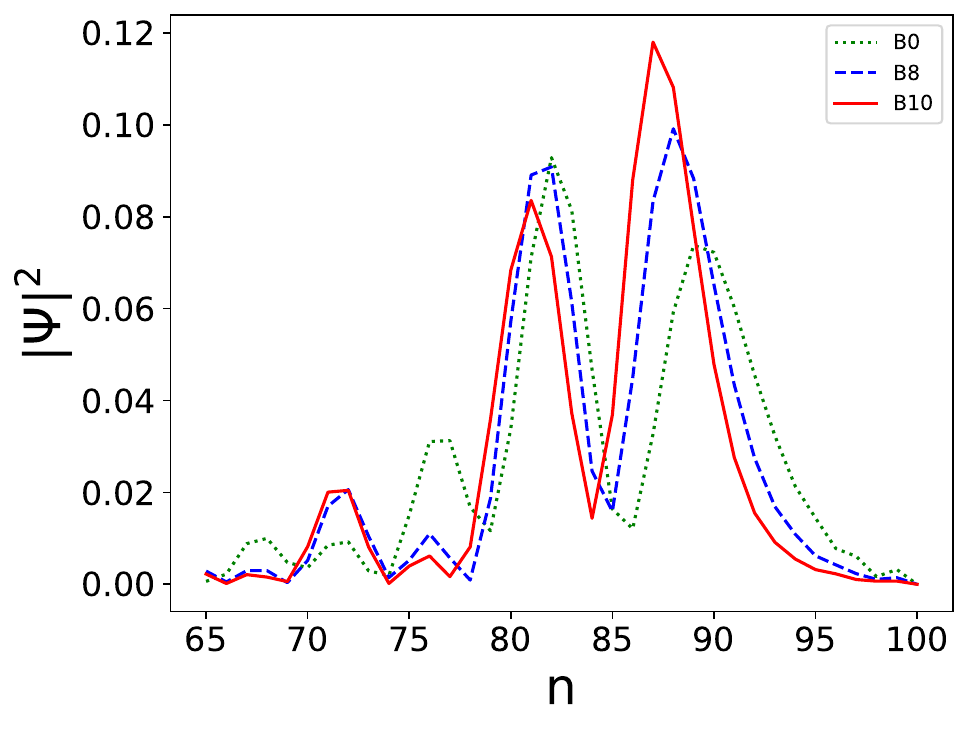}
    \includegraphics[width=60mm]{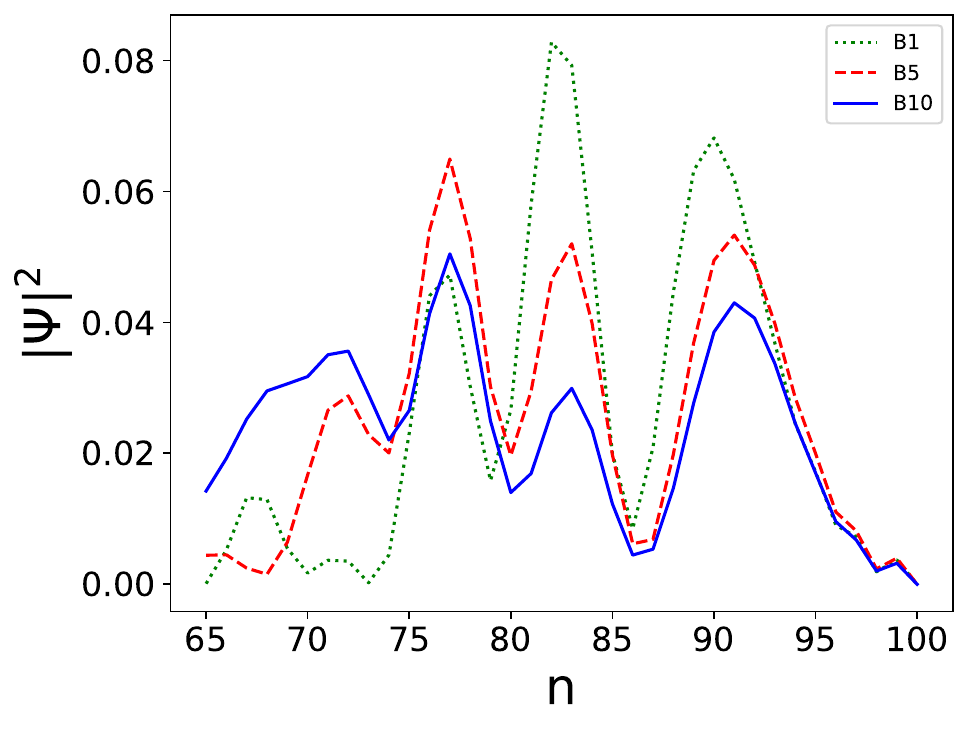}
  }
  
  \centering{ c)\hspace{60mm}d)}
  \caption {Polaron profile in a Donor-Conducting system after 48 unit of time for
    several values of the magnetic field.
    $\chi = 80$, $W=0.628$, $\sigma = 0.00646$,
    $D_d=0.1$, $J_d=0.5$, $x_d=0.3$, $v_d=0.1$, $m_d=5$.
  a) $B=1$; b) $B=10$; c) $k_y=0$; d) $L_y=5\times 10^{-9}$. 
\label{fig:CP_profiles_B_Ly}}      
\end{figure}

Figure \ref{fig:CP_profiles_B_Ly} shows the effect that the MF has
on the polaron profile and its dependence on the transverse quasi-momentum.
In the presence of a MF above the critical value, the polaron moves with a non-zero
acceleration whose value depends on the MF intensity and on the transverse quasi-momentum.
Despite this, the dynamics of the polaron remain stable even in strong MFs.

\section{Conclusions and discussion}\label{Concl}
In this paper we have studied the numerical solutions of the
coupled system of nonlinear equations
that describe the dynamics of a large polaron in low-dimensional systems in the
presence of an external magnetic field.
We have shown that the polarons in these systems are very stable
for a wide range of the parameter values
while subjected to very strong magnetic fields.
This is true even in very long molecular chains.

The polaron dynamics is affected by the presence of a
magnetic field. The characteristics of this impact depend on the field
intensity, the system parameters as well as the transverse quasi-momentum
of the polaron. Unlike for the analytical solution described in
\cite{LB-MF-SChFr}, in a discrete system there is some critical value of the
magnetic field and the polaron transverse quasi-momentum above which the
polaron can start moving when it’s initial
velocity is zero. We have also studied the dynamics of the polaron
after it was boosted and showed that in
relatively strong magnetic fields the polaron is accelerated
confirming the results of our previous analytical study \cite{LB-MF-SChFr}.

Our study shows that when the polaron is broader, one requires
a larger boost velocity to make the polaron move. This can be explained by the
fact that as a broader polaron spans over more lattice sites, the lattice phonons
of the polaron have a larger kinetic energy.

These results were obtained for a large range of system parameters,
characteristic of the Amide-I-type polaron
and electron polarons in polypeptide macromolecules, as well as for electrons
in conducting polymers, like polypyrrole and polythiophene. We have considered
magnetic fields up to 10 T as the most typical magnetic fields existing in
laboratory or used in various appliances.

We have also studied the impact of magnetic fields
on the dynamics of polarons in Donor-Molecular Chain complexes
where an electron, initially fully localised at the donor, tunnels to the chain
and generates, after a finite time, a polaron on the chain.
We have also shown that the localised electron generated by such Donor complex,
for the studied parameter values, is made of a few polarons travelling at
slightly different velocities but
leading to a highly efficient electron transport in very long chains of up
to 100 monomer units. This transport was only slightly affected even by the
stronger magnetic fields, allowing us to conclude that large soliton-like
polarons provide a highly effective long-range electron transport even in
the presence of strong external magnetic fields.

\section{Acknowledgements}
LB acknowledges grant 2025.07/0335 of the Fundamental Research Foundation of Ukraine and the support
of the INI-LMS Rebuild Ukraine Scheme at the Department of Mathematical Sciences of the University of
Durham (UK).

\bibliographystyle{unsrt}
\bibliography{Polaron-in-MF}

@Article{Ing,
author={Ing, Nicole L.
and El-Naggar, Mohamed Y.
and Hochbaum, Allon I.},
title={Going the Distance: Long-Range Conductivity in Protein and Peptide Bioelectronic Materials},
journal={The Journal of Physical Chemistry B},
year={2018},
month={Nov},
day={21},
publisher={American Chemical Society},
volume={122},
number={46},
pages={10403-10423},
issn={1520-6106},
doi={10.1021/acs.jpcb.8b07431},
url={https://doi.org/10.1021/acs.jpcb.8b07431}
}

@Article{Amdursky,
author ="Amdursky, Nadav",
title  ="Enhanced solid-state electron transport via tryptophan containing peptide networks",
journal  ="Phys. Chem. Chem. Phys.",
year  ="2013",
volume  ="15",
issue  ="32",
pages  ="13479-13482",
publisher  ="The Royal Society of Chemistry",
doi  ="10.1039/C3CP51748A",
url  ="http://dx.doi.org/10.1039/C3CP51748A"}

@article{Champion,
author = {Zhu, Yan and Champion, Richard and Jenekhe, Samson},
year = {2006},
month = {11},
pages = {},
title = {Conjugated Donor-Acceptor Copolymer Semiconductors with Large Intramolecular Charge Transfer: Synthesis, Optical Properties, Electrochemistry, and Field Effect Carrier Mobility of Thienopyrazine-Based Copolymers},
volume = {39},
journal = {Macromolecules},
doi = {10.1021/ma061861g}
}

@article{Ban,
  title={Emerging low-dimensional materials for nanoelectromechanical systems resonators},
  author={Si Young Ban and Xuchen Nie and Zhihao Lei and Jiabao Yi and Ajayan Vinu and Yang Bao and Yanpeng Liu},
  journal={Materials Research Letters},
  year={2022},
  volume={11},
  pages={21 - 52},
  url={https://api.semanticscholar.org/CorpusID:252428292}
}

@article{Karunasena,
author = {Karunasena, Chamikara and Thurston, Jonathan and Chaney, Thomas and Li, Hong and Risko, Chad and Coropceanu, Veaceslav and Toney, Michael and Bredas, Jean‐Luc},
year = {2025},
month = {01},
pages = {},
title = {Polymorphs of the n‐Type Polymer P(NDI2OD‐T2): A Comprehensive Description of the Impact of Processing on Crystalline Morphology and Charge Transport},
volume = {35},
journal = {Advanced Functional Materials},
doi = {10.1002/adfm.202422156}
}

@article{Jia,
title = {14.4% efficiency all-polymer solar cell with broad absorption and low energy loss enabled by a novel polymer acceptor},
journal = {Nano Energy},
volume = {72},
pages = {104718},
year = {2020},
issn = {2211-2855},
doi = {https://doi.org/10.1016/j.nanoen.2020.104718},
url = {https://www.sciencedirect.com/science/article/pii/S2211285520302755},
author = {Tao Jia and Jiabin Zhang and Wenkai Zhong and Yuanying Liang and Kai Zhang and Sheng Dong and Lei Ying and Feng Liu and Xiaohui Wang and Fei Huang and Yong Cao}
}

@article{Bhat,
author = {Bhattacharyya, Dhiman and Howden, Rachel M. and Borrelli, David C. and Gleason, Karen K.},
title = {Vapor phase oxidative synthesis of conjugated polymers and applications},
journal = {Journal of Polymer Science Part B: Polymer Physics},
volume = {50},
number = {19},
pages = {1329-1351},
doi = {https://doi.org/10.1002/polb.23138},
url = {https://onlinelibrary.wiley.com/doi/abs/10.1002/polb.23138},
eprint = {https://onlinelibrary.wiley.com/doi/pdf/10.1002/polb.23138},
year = {2012}
}

@article{Poudel,
author = {Bed Poudel  and Qing Hao  and Yi Ma  and Yucheng Lan  and Austin Minnich  and Bo Yu  and Xiao Yan  and Dezhi Wang  and Andrew Muto  and Daryoosh Vashaee  and Xiaoyuan Chen  and Junming Liu  and Mildred S. Dresselhaus  and Gang Chen  and Zhifeng Ren },
title = {High-Thermoelectric Performance of Nanostructured Bismuth Antimony Telluride Bulk Alloys},
journal = {Science},
volume = {320},
number = {5876},
pages = {634-638},
year = {2008},
doi = {10.1126/science.1156446},
URL = {https://www.science.org/doi/abs/10.1126/science.1156446},
eprint = {https://www.science.org/doi/pdf/10.1126/science.1156446}}

@article{Huang,
title = {Enhancing hydrogen storage properties of MgH2 through addition of Ni/CoMoO4 nanorods},
journal = {Materials Today Energy},
volume = {19},
pages = {100613},
year = {2021},
issn = {2468-6069},
doi = {https://doi.org/10.1016/j.mtener.2020.100613},
url = {https://www.sciencedirect.com/science/article/pii/S246860692030232X},
author = {T. Huang and X. Huang and C. Hu and J. Wang and H. Liu and Z. Ma and J. Zou and W. Ding}
}

@article{Chang-solar-cells,
  title={Organic solar cells based on small molecule donor and polymer acceptor},
  author={Xu, Wanru and Chang, Yilin and Zhu, Xiangwei and Wei, Zhenhua and Zhang, Xiaoli and Sun, Xiangnan and Lu, Kun and Wei, Zhixiang},
  journal={Chinese Chemical Letters},
  volume={33},
  number={1},
  pages={123--132},
  year={2022},
  publisher={Elsevier}
}

@Article{Venka,
author={Venkatasubramanian, Rama
and Siivola, Edward
and Colpitts, Thomas
and O'Quinn, Brooks},
title={Thin-film thermoelectric devices with high room-temperature figures of merit},
journal={Nature},
year={2001},
month={Oct},
day={01},
volume={413},
number={6856},
pages={597-602},
issn={1476-4687},
doi={10.1038/35098012},
url={https://doi.org/10.1038/35098012}
}

@article{Ugur,
author = {Ugur, Asli and Katmis, Ferhat and Li, Mingda and Wu, Lijun and Zhu, Yimei and Varanasi, Kripa K. and Gleason, Karen K.},
title = {Low-Dimensional Conduction Mechanisms in Highly Conductive and Transparent Conjugated Polymers},
journal = {Advanced Materials},
volume = {27},
number = {31},
pages = {4604-4610},
doi = {https://doi.org/10.1002/adma.201502340},
url = {https://advanced.onlinelibrary.wiley.com/doi/abs/10.1002/adma.201502340},
eprint = {https://advanced.onlinelibrary.wiley.com/doi/pdf/10.1002/adma.201502340},
year = {2015}
}

@book{Davydov,
  title     = "Solitons in Molecular Systems",
  author    = "Davyvov, A. S.",
  year      = 1991,
  publisher = "Reidel",
  address   = "Dordrecht"
}

@article{BPZ-DChA,
  title = {Donor-acceptor electron transport mediated by solitons},
  author = {Brizhik, L. S. and Piette, B. M. A. G. and Zakrzewski, W. J.},
  journal = {Phys. Rev. E},
  volume = {90},
  issue = {5},
  pages = {052915},
  numpages = {12},
  year = {2014},
  month = {Nov},
  publisher = {American Physical Society},
  doi = {10.1103/PhysRevE.90.052915},
  url = {https://link.aps.org/doi/10.1103/PhysRevE.90.052915}
}

@inbook{LB-Dav-periodic,
author = {Brizhik, Larissa},
year = {2017},
month = {04},
pages = {215-232},
title = {Electron Correlations in Molecular Chains},
isbn = {978-3-319-53663-7},
doi = {10.1007/978-3-319-53664-4_15}
}

@article{Sirringhaus,
       author = {{Sirringhaus}, H. and {Brown}, P.~J. and {Friend}, R.~H. and {Nielsen}, M.~M. and {Bechgaard}, K. and {Langeveld-Voss}, B.~M.~W. and {Spiering}, A.~J.~H. and {Janssen}, R.~A.~J. and {Meijer}, E.~W. and {Herwig}, P. and {de Leeuw}, D.~M.},
        title = "{Two-dimensional charge transport in self-organized, high-mobility conjugated polymers}",
      journal = {Nature},
         year = 1999,
        month = oct,
       volume = {401},
       number = {6754},
        pages = {685-688},
          doi = {10.1038/44359},
       adsurl = {https://ui.adsabs.harvard.edu/abs/1999Natur.401..685S}
}

@article{LB-MF-SChFr,
       author = {{Brizhik}, Larissa},
        title = "{Dynamics of the Davydov's soliton in external oscillating magnetic field}",
      journal = {Chaos Solitons and Fractals},
         year = 2024,
        month = oct,
       volume = {187},
          eid = {115459},
        pages = {115459},
          doi = {10.1016/j.chaos.2024.115459},
archivePrefix = {arXiv},
       eprint = {2402.09172},
 primaryClass = {cond-mat.soft},
       adsurl = {https://ui.adsabs.harvard.edu/abs/2024CSF...18715459B}
}

@article{BEC-Peierls,
  title = {Soliton dynamics and Peierls-Nabarro barrier in a discrete molecular chain},
  author = {Brizhik, Larissa and Eremko, Alexander and Cruzeiro-Hansson, Leonor and Olkhovska, Yulia},
  journal = {Phys. Rev. B},
  volume = {61},
  issue = {2},
  pages = {1129--1141},
  numpages = {0},
  year = {2000},
  month = {Jan},
  publisher = {American Physical Society},
  doi = {10.1103/PhysRevB.61.1129},
  url = {https://link.aps.org/doi/10.1103/PhysRevB.61.1129}
}

@article{Gulacsi,
author = {M. Gulácsi and M.A.M. El-Mansy and Z. Gulácsi},
title = {Electron-phonon interactions in conducting polymers},
journal = {Philosophical Magazine Letters},
volume = {96},
number = {2},
pages = {67--75},
year = {2016},
publisher = {Taylor \& Francis},
doi = {10.1080/09500839.2016.1150611},
URL = {https://doi.org/10.1080/09500839.2016.1150611},
eprint = {https://doi.org/10.1080/09500839.2016.1150611}
}

@Article{Le,
AUTHOR = {Le, Thanh-Hai and Kim, Yukyung and Yoon, Hyeonseok},
TITLE = {Electrical and Electrochemical Properties of Conducting Polymers},
JOURNAL = {Polymers},
VOLUME = {9},
YEAR = {2017},
NUMBER = {4},
ARTICLE-NUMBER = {150},
URL = {https://www.mdpi.com/2073-4360/9/4/150},
PubMedID = {30970829},
ISSN = {2073-4360},
DOI = {10.3390/polym9040150}
}

@Article{Aziz,
author ="Abdel Aziz, Ilaria and Tullii, Gabriele and Antognazza, Maria Rosa and Criado-Gonzalez, Miryam",
title  ="Poly(3-hexylthiophene) as a versatile semiconducting polymer for cutting-edge bioelectronics",
journal  ="Mater. Horiz.",
year  ="2025",
volume  ="12",
issue  ="15",
pages  ="5570-5593",
publisher  ="The Royal Society of Chemistry",
doi  ="10.1039/D5MH00096C",
url  ="http://dx.doi.org/10.1039/D5MH00096C"}

\end{document}